\documentclass[twocolumn,showpacs,superscriptaddress,longbibliography,nofootinbib,pra]{revtex4-1}
\usepackage[utf8]{inputenc}
\usepackage{calrsfs}
\usepackage{graphicx}
\usepackage{textcomp}
\usepackage{soul}
\usepackage{amsmath,amssymb}
\usepackage{color}
\usepackage{braket}
\usepackage{xcolor}
\usepackage{float}
\usepackage{physics}
\usepackage{soul}
\usepackage{graphicx}
\usepackage{caption}
\usepackage{amsmath, amssymb}
\usepackage{bbold}
\usepackage{makecell}
\usepackage{tikz,pgfplots}
\usetikzlibrary{decorations.pathreplacing,arrows.meta}
\usepackage{mathtools}
\usepackage{multirow}
\usepackage{subcaption}
\usepackage{url}
\usepackage[colorlinks=true, urlcolor=blue, citecolor=blue, linkcolor=blue]{hyperref}
\begin{document}
\title{SYK model based $\beta$ regime dependent two-qubit dynamical wormhole-inspired teleportation protocol simulation}
\author{Sudhanva Joshi }
\email[]{sudhanvajoshi.rs.phy24@itbhu.ac.in}
\affiliation{Department of Physics, Indian Institute of Technology (Banaras Hindu University), Varanasi - 221005, India}

\author{Sunil Kumar Mishra}
\email[]{sunilkm.app@iitbhu.ac.in}
\affiliation{Department of Physics, Indian Institute of Technology (Banaras Hindu University), Varanasi - 221005, India}

\begin{abstract}

We implement the Wormhole-Inspired Teleportation Protocol (WITP) in a pair of coupled Sachdev-Ye-Kitaev (SYK) models prepared in a thermofield-double state, forming a quantum analog of a traversable wormhole. By varying parameters (temperature $\beta$, coupling strength $g$, insertion site and traversal time $t$) we compare the teleportation fidelity against an analogous protocol using a transverse-field Ising model. We find that the chaotic SYK system consistently yields higher teleportation fidelity than the TFIM model, reflecting the SYK Hamiltonian’s pronounced many-body chaos. These enhanced fidelities arise from the SYK’s effectively random-matrix dynamics, which improve the coherent information transfer through the wormhole channel. Unlike prior single-qubit benchmarks based on $\ket{0}$ inputs, the present work defines and evaluates a genuinely quantum-state fidelity for a maximally entangled two-qubit Bell input, using a Pauli-stabilizer formalism that captures entanglement-phase coherence. Our central result is achieved by teleporting a maximally entangled two-qubit Bell state through the wormhole. We introduce a Pauli-stabilizer fidelity measure for the two-qubit message and demonstrate that the Bell-state protocol produces a substantial fidelity boost compared to single-qubit teleportation. Furthermore, we examine the time-resolved fidelity for both single-qubit and two-qubit messages, revealing distinct fluctuation patterns that deepen the understanding of dynamical many-body teleportation processes. Finally, we present an argument that our Bell-state WITP simulations provide a concrete numerical testbed for aspects of the ER = EPR conjecture, by mapping entanglement structure and thermal/coupling dependence to traversability diagnostics in an emergent wormhole geometry. \\


\end{abstract}

\maketitle
\section{Introduction}
Ever since the pioneering Maldacena paper \cite{maldacena1999large}, there have been large-scale efforts to understand the implications of AdS/CFT correspondence and the holographic duality, with much of the development lately having been on simulating specific effects or sections of holographic duality using a quantum computer or high-performance clusters. Various works like \cite{brown2019quantum,alba2019quantum,jafferis2022traversable,Shapoval2023towardsquantum} delve into quantum gravity and holographic duality from a perspective of spin chains of qubits and Majorana fermions\cite{qi2020coupled}. They posit a quantum dual to the gravitational picture of Anti-de Sitter (AdS) Space as an entangled system of two wormholes. Most works take advantage of a novel concept known as traversable wormholes. Traversable wormholes are curved spacetime geometries with two asymptotically AdS regions in the bulk connected by a wormhole with a CFT living on both boundaries. Although going by General Relativity, traversable wormholes are inaccessible due to the Average Null Energy Condition (ANEC) \cite{hartman2017averaged}, Gao {\it et al.}\cite{gao2017traversable} hypothesized that wormholes can indeed be made traversable by inducing a coupling between the boundaries of AdS/CFT wormhole geometry. The AdS bulk dynamics and the deformation Hamiltonian govern it. There is maximal entanglement between the two boundary CFTs. Traversable wormholes allow direct information propagation when the coupling is appropriately tuned. As demonstrated in \cite{brown2023quantum} and \cite{schuster2022many}, the established protocol uses the operator SIZE winding mechanism to allow teleportation to occur successfully. They first take a kicked TFIM Hamiltonian \cite{bertini2018exact,bertini2019entanglement} and describe its time evolution in two parts, followed by coupling the sides via a coupling constant $g$ to unwind the phases, followed by tracing the qubit out to study the fidelity. Their findings reveal the existence of a teleportation window for particular values of the coupling constant $g$ when there is an alignment of phases.


\begin{figure*}
\centering
    \begin{tikzpicture} [scale=1.0]

\draw[very thick] (-4,0) node[above] {$\ket{\psi}_{initial}$}  -- (-2.125,0);
\draw[very thick] (-1.625,0) -- (4,0);
\draw[very thick]  (-4,-0.5) -- (-2.125,-0.5);
\draw[very thick]  (-1.625,-0.5) -- (4,-0.5);
\draw[very thick] (-4,-1) -- (4,-1);
\draw[very thick]  (-4,-1.5) -- (4,-1.5);
\draw[very thick] (-4,-2) -- (4,-2);
\draw[very thick] (-4,-2.5) -- (4,-2.5);
\draw[very thick] (-4,-3) -- (4,-3)node[below right] {$\ket{\psi}_{final}$};

\node[draw=black, fill=white, rounded corners=5pt, minimum width=1cm, minimum height=1.5cm, align=center] at (-3,-1) {$e^{iH_{L}t}$};
\node[draw=black, fill=white, rounded corners=5pt, minimum width=1cm, minimum height=1.5cm, align=center] at (-0.625,-1) {$e^{-iH_{L}t}$};
\node[draw=black, fill=white, rounded corners=5pt, minimum width=1cm, minimum height=1.5cm, align=center] at (3,-2.5) {$e^{-iH_{R}t}$};
\node[draw=black, fill=white, rounded corners=5pt, minimum width=1cm, minimum height=2.25cm, align=center] at (1,-1.75) {$e^{igV}$};

  \draw (-2.125,0) -- (-1.625,-0.5);
  \draw (-2.125,-0.5) -- (-1.625,0);
 \draw[very thick] (-4,-0.5) arc[start angle=-270, end angle=-90, radius=1.25 cm];
 \draw[very thick] (-4,-1) arc[start angle=-270, end angle=-90, radius=0.75 cm];
  \draw[very thick] (-4,-1.5) arc[start angle=-270, end angle=-90, radius=0.25 cm];

  \node[fill=red, fill opacity=0.3, text opacity=1, draw=blue, rounded corners, minimum width=2cm, minimum height=1.5cm, align=center] at (-5,-0.875)
    {\Huge L};

  \node[fill=red, fill opacity=0.3, text opacity=1, draw=blue, rounded corners, minimum width=2cm, minimum height=1.5cm, align=center] at (-5,-2.625)
    {\Huge R};

\end{tikzpicture}

\caption{Quantum circuit representation of Wormhole-Inspired teleportation protocol. L and R are the two throats of a geometric wormhole initialized in the TFD state. On the left side, we have negative time evolution followed by the swapping in of the message qubit. Subsequently, there is time evolution through the L side first and coupling with the R side, followed by its traversal throughout the system, only to be traced out from the R throat.}
 \label{QC}
\end{figure*}

Moving towards the SYK model, it is a known fact that the SYK Hamiltonian is highly chaotic \cite{maldacena2016bound} with a high degree of instant saturation in out-of-time-order correlation value \cite{garcia2022out}. Gao-Jafferies-Wall \cite{gao2021traversable} demonstrated the Wormhole-inspired teleportation protocol (WITP) using the SYK Hamiltonian. Their protocol setup includes two copies of the coupled SYK system in a TFD state, representing a dual to wormhole \cite{maldacena2003eternal}. After swapping in the message qubit, an appropriately tuned coupling term composed of complex Majorana operators is used to enact traversability of the wormhole via negative energy shockwave \cite{hirano2019information}, followed by a trace-out operation to assess Mutual information $\mathit{I(L:R)}$ and two point correlation function for left and right SYK copies. Our approach is to analyze the protocol within the framework of a quantum circuit and compute the fidelity values by performing a systematic three-parameter exploration of teleportation quality - sweeping coupling strength ($g$), traversal time($t$) and temperature ($\beta$). Fidelity measure gives a more fine-grained operational benchmark for assessing the quality of quantum teleportation. While a channel can have a high $\mathit{I(L:R)}$ value due to shared noise and classical correlations, it may fail to teleport a qubit with high fidelity. Essentially, by computing fidelity, we can exploit a channel's functional quantum teleportation capacity. This elucidates how well a qubit's information would be teleported across the channel. 

This protocol works robustly in the Sparsed SYK model as well \cite{jafferis2022traversable,caceres2021sparse,garcia2021sparse}. The mechanism involved in the SYK-based protocol is the same as the TFIM-based protocol. Teleportation relies on Size winding where the phase of the size distribution mimics the particle's momentum, enabling transmission through the bulk of the wormhole. The coupling $e^{igV}$ unwinds this winding, mapping the operator from one boundary to the other \cite{nezami2023quantum} illustrated in Fig.(\ref{QC}). One common theme in both models is using the TFD state in its maximally entangled form, i.e., in $\beta=0$ regime. In this paper, we will investigate the effects of using non-zero $\beta$ on the teleportation fidelity of the system. In doing so, we will compute the relationship between the physical placement of Majorana operators on qubit indices and generated random phases in the SYK system by defining overlap coefficients ($C_{nm}^{(ij)}$) dependent on the decomposition of the SWAP operator. This relationship is a complex one and has a penchant for having maximum overlap in the maximally entangled limit ($\beta=0$) and $1^{\rm st}$  qubit insertion ($\mathrm{SWAP(0,1)}$). Since $C_{nm}^{(ij)} \propto \mathcal{F}_{Z}(\beta)$, this correlates to maximum fidelity of teleportation when those conditions are met. This is because information on a swapped-in qubit quickly scrambles into many-body degrees of freedom in a chaotic SYK system. As we go from maximally entangled to maximally mixed ($\beta >>1$) TFD state, the scrambling maximizes \cite{iyoda2018scrambling}. This property of the SYK system makes the whole setup accurate for being an analog to information scrambling in an entangled pair of blackholes or a wormhole system \cite{sekino2008fast,lashkari2013towards}. \\


To have a statistically significant teleportation fidelity, the qubit's information must remain coherent until the L and the R sides are coupled with a coupling constant $g$. This weak coupling is turned ``on" at zero time. We get information transmission only in a narrow window of $g$ and $t$. TFIM-based protocol yields the same characteristic behavior. Information passes through the bulk and traverses the entire region only to appear at the R side at a suitable range of $g$. We investigate this behavior for the SYK-based protocol considering a broad range of $\beta$ values, and $\mathrm{SWAP(0,1)}$ and $\mathrm{SWAP(0,2)}$ operators. \\
Moving on to two-qubit teleportation, we teleport a Bell state ($\Phi^+$) through the wormhole and study its information dispersion characteristics. To study its fidelity, we must develop a new definition of Pauli stabilizer-based fidelity to account for a two-channel Bell state message. We then compare its fidelity values to ascertain the entanglement behavior \cite{hu2013relations, clauser1969proposed}. We aim to be consistent with the fact that WITP, in general, geometerizes quantum teleportation, with SYK dynamics producing features of Jackiw-Teitelboim gravity \cite{jackiw1985lower} and semiclassical wormhole physics \cite{gao2021traversable}. We then conclude the results of the paper in the last section.

\section{Wormhole-Inspired Teleportation via SYK Formalism}
Since Majorana fermions have their own antiparticles,
\newline
$\{c_{i},c_{j}\} = 0$ if $i\neq j$ and $\{c_{i},c_{j}^{\dagger}\}= \delta_{ij}$ which is a canonical fermionic relation \cite{kitaev2001unpaired}. Here $c_{i}$ is represented using Jordan-Wigner transformation \cite{nielsen2005fermionic}:
\begin{equation}
    c_{i}= I_{1}\otimes I_{2} \otimes\cdot\cdot\cdot \otimes I_{i-1} \otimes a_{i} \otimes Z_{i+1} \otimes \cdot \cdot\cdot\otimes Z_{n}.
\end{equation}
$a_{i}$ is the annihilation operator and $Z_{n}$ ensures correct anti commutation relation. Now we split fermions into Majorana modes:
\begin{equation}
    \gamma_{2i}= c_{i}+c_{i}^{\dagger}, \gamma_{2i+1} = i(c_{i}-c_{i}^{\dagger}) ,
\end{equation}
where the Majorana modes satisfy
\begin{equation}
    \{\gamma_{2i},\gamma_{2j}\} = 2\delta_{ij} .
\end{equation}

We then define a orthonormal operator basis $\{\mathcal{M}_{s}\}$ for the Hilbert space where each $\{\mathcal{M}_{s}\}$ is a product of Majoranas
\begin{equation}
    \mathcal{M}_{s} = \gamma_{i1}\gamma_{i2}\cdot\cdot\cdot\gamma_{ik}, S \subset\{1,2,\cdot\cdot\cdot,n\} .
\end{equation}
This includes the identity operator $\mathcal{M}_{\emptyset} = \textit{I}$. 
We now prepare a TFD state, which is a specially entangled state that can be thought of as representing two entangled SYK Hamiltonians $H_{L}$ and $H_{R}$ at a given inverse temperature $\beta$ \cite{antonini2023holographic}
\begin{equation}
    \ket{TFD} = \frac{1}{\sqrt{Z}} \sum_{i} e^{-\beta E_{i}/2} \ket{E_{i}}_{L} \otimes \ket{E_{i}}_{R} .
\end{equation}
$\textit{Z}$ is a normalization factor (partition function) and $\ket{E_{i}}$ are eigenstates of $H$ which satisfy $H\ket{E_i}=E_{i}\ket{E_i}$. The TFD state is intriguing because we can decompose it into a pair of $N$ Bell pairs in an infinite temperature regime \cite{zhu2020generation}.
\begin{equation}
    \ket{TFD} = \ket{\Phi^+}^{\otimes N} ,  \beta = 0 .
\end{equation}
We have the $q=4$ $N=6$ SYK Hamiltonian$(H_{SYK}) $ with all to all interactions which we decompose into $H_{L}$ and $H_{R}$ where  $H_{L}$ and $H_{R}$ are two Hamiltonians of L and R subsystem respectively \cite{rosenhaus2019introduction,maldacena2021syk} 
\begin{equation} \label{HL}
    H_{L} = -\frac{1}{q!}\sum_{i<j<k<l} \textit{J}_{ijkl}\gamma_{Li}\gamma_{Lj}\gamma_{Lk}\gamma_{Ll}.
\end{equation}
Similarly, for the Right side,
\begin{equation} \label{HR}
     H_{R} = -\frac{1}{q!}\sum_{i<j<k<l} \textit{J}_{ijkl}\gamma_{Ri}\gamma_{Rj}\gamma_{Rk}\gamma_{Rl}.
\end{equation}
Each Hamiltonian consists of all possible quartic interactions between Majorana modes on the respective side. Here, the indices $i,j,k,l$ run over the Majorana modes on each side, and the coupling constants $J_{ijkl}$ are drawn from a Gaussian distribution such that
\begin{equation}
    \expval{J_{ijkl}} = 0 , \expval{J_{ijkl}^{2}} = \frac{J_{ijkl}^{2}(q-1)!}{n^{q-1}} .
\end{equation}
Our simulations are performed for a system size of $N=6$ Majorana modes per side. While the holographic properties of the SYK model are formally defined in the large-$N$ limit, a detailed numerical study at $N=6$ provides a significant and physically robust contribution for several reasons. First, the Hilbert space dimension grows exponentially with $N$, making exact diagonalization computationally prohibitive for larger systems. Second, the small-$N$ regime is not merely a finite-size approximation but can be a qualitatively different physical system of interest. Indeed, $N=6$ is a critical system size, lying at the threshold where variants of the SYK model are conjectured to transition from integrable to chaotic behavior. Finally, this system size aligns with the current frontier of experimental and numerical capabilities. Recent quantum simulations of traversable wormhole dynamics have been implemented on quantum processors using comparable resource counts, such as the nine-qubit circuit for a sparsified $N \approx 7$ SYK model by \cite{jafferis2022traversable} and the $N=7$ spin simulation of the kicked Ising model by \cite{brown2023quantum}. Our choice of $N=6$ thus provides a precise, non-trivial investigation of a system large enough to exhibit many-body chaos while remaining small enough for a high-precision exact diagonalization study.
\footnote{We note that extending to $N=8$ would require $\sim 16\rm x$ more memory and $\sim 64\rm x$ more computation time for exact diagonalization, placing it beyond our current computational budget. Approximate methods sacrifice the exactness that makes N=6 results definitive benchmarks.}
\\

We then define the INSERT operator as a critical component in the wormhole protocol. Mathematically, it is responsible for encoding the message qubit into the left subsystem of the TFD state. The INSERT operator is given as 
\begin{equation} \label{Insert}
    \hat{I} = \sum_{\alpha,\beta} c_{\alpha\beta}\sigma_{\mathrm{message}}^\alpha \otimes M_{L}^{\beta} ,
\end{equation}
where the $\sigma_{\mathrm{message}}^\alpha$ involves the Pauli basis operators indexed by $\alpha \in(\textit{I,X,Y,Z})$ that act on the message qubit. The Majorana operators acting on the left subsystem $(M_{L}^{\beta})$ represent the insertion of the message into the wormhole. These are indexed by $\beta$ corresponding to combinations of Majorana operators in the left Hilbert space. Coefficients for the linear combination $ c_{\alpha\beta}$, determined by decomposing the SWAP operator in the Pauli basis.\newline
To introduce a nonlocal interaction that creates correlations between the left and right subsystems, we define the SIZE operator based on \cite{qi2019quantum} as
\begin{equation}
    \hat{\upsilon} = \sum_{i=2}^{N} c_{i}^{\dagger}c_{i},
\end{equation}
where $c_{i}$ are fermionic annihilation operators acting on the Majorana modes of the system. The coupling term $e^{ig\hat{\upsilon}}$ measures the occupation number on the right side. Mathematically, when the coupling term operates on any state $\ket{\psi}$, we get
\begin{equation}
    e^{ig\hat{\upsilon}}\ket{\psi} = \sum_{n=0}^{\infty} \dfrac{(ig)^{n}}{n!} \hat{\upsilon}^{n}\ket{\psi} .
\end{equation}
$\hat{\upsilon}$ is diagonal in the occupation number basis (eigenstates of 
SIZE operator corresponds to specific occupations) and $\ket{u_n}$ are eigenstates of $\hat{\upsilon}$ and $\lambda_{n}$ are corresponding eigenvalues.
This introduces a phase shift depending on the occupation number, effectively coupling the two sides of the TFD state \cite{roberts2018operator}. The strength of these correlations is proportional to $g$. We have
\begin{equation} \label{size}
    e^{ig\hat{\upsilon}}\ket{\psi} = e^{ig\lambda_{n}}\ket{\psi} .
\end{equation}
From Figs. (\ref{sq1}, \ref{sq2}, \ref{fig34}), we can observe a definitive periodicity in teleportation fidelity as we coarse over coupling strength $g$. This observed periodicity of the teleportation fidelity as a function of $g$ originates from the spectral properties of the SIZE operator. Since $\hat{\upsilon}$ is diagonal in the occupation number basis with integer eigenvalues $\lambda_n \in \{0,1,2,...,N-1\}$, its exponential form $e^{ig\hat{\upsilon}}$ acts as a phase rotation operator:
\begin{equation}
    e^{ig\hat{\upsilon}}\ket{\psi} = \sum_n e^{ig\lambda} \expval{u_n|\psi}\ket{u_n}.
\end{equation}
The vectors $\ket{u_n} \equiv \ket{n_2,n_3,...,n_N}, n\in\{0,1\}$ denote the eigenstates of the SIZE operator, which satisfy the condition $\hat{\upsilon}\ket{u_n} = \lambda\ket{u_n}$. Each eigenstate corresponds to a definite pattern of occupied and unoccupied fermionic modes, and its eigenvalue is the total number of occupied modes:
\begin{equation}
    \lambda_n = \sum_{i=2}^{N} n_i
\end{equation} 
Our initial state in the Z basis is defined as 
\begin{equation} \label{ini=msgtensortfd}
    \ket{\psi}_{\mathrm{initial}} = \ket{\mathrm{message}} \otimes \ket{TFD}.
\end{equation}
To Provide the most general analysis and to address the distinction between classical and quantum fidelity, we begin by considering the message qubit initialized in an arbitrary superposition state $\ket{\psi}_{\rm message} = \alpha\ket{0}+\beta\ket{1}$, where $|\alpha|^{2}+|\beta|^{2}=1$. A full fidelity analysis for this arbitrary input state, which we perform in detail in Appendix \ref{appendix2}, separates the problem into two distinct components, namely the `classical' probability of successfully transmitting the basis state(the diagonal terms of output density matrix) and the `quantum' coherence of superposition(off-diagonal terms). \\
As we demonstrate both analytically and numerically in the Appendix, the chaotic and thermal dynamics of the SYK model actively suppress the off-diagonal coherence terms, causing the quantum part of the signal to decohere. Our simulations (Fig.~(\ref{neofidelity})) explicitly show that the fidelity averaged over random input states ($\overline{\mathfrak{F}}_{\alpha\beta}$) exhibits the same qualitative dependence on coupling $g$ and temperature $\beta$, but quantitatively, a scaled-down echo due to decoherence of off-diagonal terms as compared to fidelity of the basis state $\ket{0}$ alone.\\
This confirms that the basis-state fidelity $\mathfrak{F}_z$ in Eq.~(\ref{sqfide}), is not merely a ``classical" benchmark but serves as a faithful and robust proxy for the full, averaged quantum channel performance. Therefore, for clarity and to follow the standard methodology in foundational works \cite{{Shapoval2023towardsquantum},{brown2023quantum},{nezami2023quantum}}, we initialize the message in the $\ket{0}$ basis.
The message qubit is now initialized in the $\ket{0}$ state, having an eigenstate of $\sigma_{z}$ with the eigenvalue being $+1$.
Before applying the $\hat{I}$ operator, we evolve the system backward in time to avoid the evolution of our state straight into the singularity, as can be intuitively seen from Fig.~(\ref{WITP bulk boundary geometry}). 

\begin{figure}
\centering

\begin{tikzpicture}[scale=1]

\fill[blue!10] (-3,3) -- (0,0) -- (0,0) -- (-3,-3) -- cycle;
\draw[thick,blue] (-3,3) -- (0,0) ;
\draw[thick,blue] (-3,-3) -- (0,0) node[midway,left] {Left CFT};

\fill[red!10] (3,3) -- (0,0) -- (0,0) -- (3,-3) -- cycle;
\draw[thick,red] (3,3) -- (0,0) ;
\draw[thick,red] (3,-3) -- (0,0) node[midway,right] {Right CFT};

\draw[thick,dashed] (-1,1) to[out=0,in=180] (1,1) node[midway,above] {AdS Bulk};
\draw[thick,dashed] (-1,-1) to[out=0,in=180] (1,-1);

\draw[->] (-3,0) -- (3,0) node[below right] {$x$};
\draw[->] (0,-3) -- (0,3) ;

\draw (-3,0) -- (3,0);
\draw (-3,-3) .. controls (0,0) .. (3,-3);
\draw (-3,3) .. controls (-1,0) and (1,0) .. (3,3);


\node at (-2.5,2.5) [blue] {Left throat};
\node at (2.5,2.5) [red] {Right throat};
\node at (-3.3,0)[black] {t=0 $\bullet$};
\draw[->,thick,black](-3.05,-0.1) -- (-3.05,-1.9) node[midway,left] {$e^{iH_{L}t}$};
\draw[->,thick,black](3.05,0.1) -- (3.05,1.9) node[midway,right] {$e^{-iH_{R}t}$};
\node at (-3.47,-3)[black] {t=-$\infty$ $\bullet$};
\node at (-3.47,3)[black] {t=$\infty$ $\bullet$};
\draw[thick,decorate,decoration={snake}] (-3,3) -- (3,3) node[midway,above] {Singularity};
\draw[thick,decorate,decoration={snake}] (-3,-3) -- (3,-3) node[midway, below] {Singularity};

\draw[orange,thick,decorate,decoration={snake,amplitude=0.9mm,segment length=5mm}] (-3,0) -- (0,3); \node at (0,2.4)[orange]{Coupling};
\draw[orange,thick,decorate,decoration={snake,amplitude=0.9mm,segment length=5mm}] (3,0) -- (0,3);   
\draw[very thick, black](1,2) -- (2,1) ;
\draw[->,very thick,black] (2,1) -- (3,2) node[right] {$\ket{\psi}_{final}$} ;
\draw[thick,decorate,decoration={snake,amplitude=0.5mm,segment length=5mm}] (-3,-2) -- (1,2) ; \node at (-3.1,-2.1) {$\ket{\psi}_{initial}$};
    
\end{tikzpicture}

 \caption{Wormhole-inspired teleportation protocol in AdS/CFT Geometry picture. In order to avoid the transmission of our message straight into the singularity, from L throat, we start with a negative time evolution, followed by the swapping in of the message and positive time evolution. The coupling introduces a ``shockwave" which propels the message out through the R throat of the wormhole.}
 \label{WITP bulk boundary geometry}
\end{figure}

 From \cite{bhattacharyya2022quantum}, we define time evolution in bulk as a wormhole unitary transformation, which simulates the interaction between the left and right Hamiltonians with a coupling term $g$ along with temporal evolution of the state under this transformation as 
\begin{equation} \label{wunitary}
    U_{\mathrm{wormhole}}  = e^{-iH_{R}t}e^{ig\hat{\upsilon}}e^{-iH_{L}t}\hat{I}e^{iH_{L}t} .
\end{equation}
Our final state, based on the unitary time evolution, now becomes
\begin{equation}
      \ket{\psi}_{\mathrm{final}} = U_{\mathrm{wormhole}} \ket{\psi}_{initial} .
\end{equation}
 In the full Wormhole Unitary Eq.~(\ref{wunitary}), the operator $e^{ig\hat{\upsilon}}$ introduces eigen-value dependent phase factor demonstrated in Eq.~(\ref{size})The teleportation fidelity contains interference terms of the form $e^{ig(\lambda_n-\lambda_m)}$. This happens because all differences $(\lambda_n-\lambda_m)$ are integers, every such term is exactly $2\pi$-periodic in $g$,i.e.,
\begin{equation}
    e^{i(g+2\pi)(\lambda_n-\lambda_m)} = e^{ig(\lambda_n-\lambda_m)}.
\end{equation}
Consequently, the overall fidelity function satisfies $\mathcal{F}(g+2\pi) = \mathcal{F}(g)$, yielding the observed recurrence of high-fidelity teleportation windows at intervals of approximately $2\pi$ in $g$.

The Pauli operator Z, which refers to the Pauli-$Z$ acting on the message qubit of the right subsystem as $Z_R$ and the one acting on the left subsystem as $Z_L$. The expectation value on the left i.e., $\expval{Z}_L$ will be +1 (as $\ket{0}$ is the message state \cite{brown2019quantum}). Operationally, $\expval{Z}_R$ quantifies how well the teleported qubit on the right reproduces the original input encoded on the left. For a single qubit reference case, i.e.,$\ket{0}$ as an input state, $\expval{Z}_R$ simply measures recovery of logical basis state and corresponds to a classical benchmark fidelity which we denote as \cite{brown2019quantum,Shapoval2023towardsquantum}:
\begin{equation} \label{sqfide}
   \mathcal{F}_{z} = \expval{Z}_{R} = \bra{\psi_{\mathrm{final}}}Z\ket{\psi_{\mathrm{final}}} .
\end{equation}

To maximize the $\expval{Z}_{R}$ value, we compute the averaged state with an array of $g$ values. Doing so gives us optimal $g$ against maximum $\expval{Z}_{R}$. 
Based on \cite{nezami2023quantum,brown2019quantum,schuster2022many} we assert that the expectation value $\expval{Z}_{R}$ measures the message ``passing" or teleportation capability of wormhole transformation, serving as a metric of how effectively information is preserved in the evolved state.

\section{Variational temperatures in the TFD state}
\subsection{Zero and finite $\beta $ regimes}
In the infinite temperature regime, as we know, the TFD state decomposes into $N$ Bell pairs \cite{cottrell2019build}. It is a maximally entangled state between the left and right subsystems of the SYK model
\begin{equation}
    \ket{TFD}_{\beta = 0} = \sum_{n}\mathfrak{c}_{n}\ket{n}_{L}\otimes\ket{n}_{R} ,
\end{equation}
$\ket{n}_{L}$ and $\ket{n}_{R}$ are eigenstates of the SYK Hamiltonian on the left and right subsystems, and $\mathfrak{c_{n}}$ is the coefficient. From Eq.(\ref{ini=msgtensortfd}), we know that the initial state for the system is an overlap of the message and TFD state.
After applying $\hat{I}$ and coupling, we get a phase to each eigenstate of $\hat{\upsilon}$. Thus, we get a state in the bulk space \cite{cardy1991bulk}
\begin{equation}
  \ket{\psi}_{\mathrm{bulk}} = \sum_{n}\mathfrak{c_{n}}e^{ig\lambda_{n}}\ket{n}_{\mathrm{message}}\otimes\ket{0}_{L}\otimes\ket{n}_{R} .
\end{equation}
The time evolution operator modifies the right subsystem states as
\begin{equation}
    e^{-iH_{R}t}\ket{n}_{R} = \sum_{m}\alpha_{nm}\ket{m}_{R},
\end{equation}
where $\alpha_{nm}$ are time-evolved coefficients.
The final state of the system after further simplification can be written as
\begin{equation}
    \ket{\psi}_{\mathrm{final}}^{\beta=0} =\sum_{n,m}\Tilde{c}_{nm}\ket{n}_{\mathrm{message}}\otimes\ket{0}_{L}\otimes\ket{m}_{R} ,
\end{equation}
where $\Tilde{c}_{nm}= \mathfrak{c_{n}}e^{ig\lambda_{n}}\alpha_{nm} $. \\

As we go from the infinite temperature regime to the finite one, the TFD state no longer consists of ad-hoc $N$ Bell pairs. There is an addition of Boltzmann weights, which dictate the energy level associated with the states present in the ensemble, which is denoted as
\begin{equation}
    \ket{TFD}_{\beta\neq0} = \frac{1}{\sqrt{Z}}\sum_{n}e^{-\beta E_{n}/2}\ket{n}_{L}\otimes\ket{n}_{R} .
\end{equation}
Here, $Z=\sum_{n}e^{-\beta E_{n}}$ is a partition function that normalizes the state. This replaces coefficients $\mathfrak{c_{n}} = \frac{1}{\sqrt{N}}$ in $\beta = 0$ to $\mathfrak{c_{n}} = \frac{e^{-\beta E_{n}/2}}{\sqrt{Z}}$ such that
\begin{equation}
    \ket{\psi}_{\mathrm{initial}} = \ket{0}_{\mathrm{message}} \otimes \frac{1}{\sqrt{Z}}\sum_{n}e^{-\beta E_{n}/2}\ket{n}_{L}\otimes\ket{n}_{R} .
\end{equation}
After evolving the state through $\hat{I}$ and $\hat{\upsilon}$ operations to introduce a phase for each eigenstate, we get
\begin{equation}
    \ket{\psi}_{\mathrm{bulk}} = \frac{1}{\sqrt{Z}}\sum_{n}e^{-\beta E_{n}/2}e^{ig\lambda_{n}}\ket{n}_{\mathrm{message}}\otimes\ket{0}_{L}\otimes\ket{n}_{R} .
\end{equation}
Evolving the state through the right side and defining $\alpha_{nm} = \bra{m}e^{-H_{R}t}\ket{n}$, the final state in this regime will read as

\begin{equation}
    \ket{\psi}_{\mathrm{final}}^{\beta \neq 0} = \sum_{n,m}\Tilde{d}_{nm}\ket{n}_{\mathrm{message}}\otimes\ket{0}_{L}\otimes\ket{m}_{R} ,
\end{equation}
where $\Tilde{d}_{nm} = \frac{1}{\sqrt{Z}}e^{-\beta E_{n}/2}e^{ig\lambda_n}\alpha_{nm}$.
Comparing the $\beta=0$ and $\beta \neq 0$ regimes, it is evident that inclusion of the Boltzmann weight factor in $\Tilde{d}_{nm}$ results in the reduction of overlap between the two coupled sides.
\subsection{Fidelity dependence on $\beta$ regime and Overlap Coefficients} \label{3b}
In the above subsection, we show how the Boltzmann weight is added to the final state in $\beta\neq0$ regime in the form of $\Tilde{d}_{nm}$. This results in a change in the value of fidelity($\mathcal{F}(\beta)$) owing to 
\begin{equation}
    \expval{Z}_{R}^{\beta=0} \propto Tr(\rho_{\mathrm{final}}Z),
\end{equation}
\begin{equation}
     \expval{Z}_{R}^{\beta\neq0} \propto Tr(\rho_{\mathrm{final}}e^{-\beta H}Z).
\end{equation}
The TFD state equally weights all eigenstates at $\beta=0$, leading to maximal entanglement between the left and right subsystems. This strong entanglement causes rapid decoherence when the message qubit interacts with one subsystem and is swapped to the other. Nevertheless, at non-zero inverse temperatures($\beta\neq0$), the TFD state is dominated by lower-energy eigenstates due to the Boltzmann factor, leading to lowered fidelity on the right.\par
The INSERT operator is constructed using a SWAP operation between the message qubit and the left subsystem. Mathematically, $\mathrm{SWAP}=\sum_{\alpha,\beta}c_{\alpha\beta}\sigma_{\mathrm{message}}^{\alpha}\otimes\sigma_{\mathrm{left}}^{\beta}$ where $\sigma^{\alpha}$ and $\sigma^{\beta}$ are Pauli basis operators. We then map Pauli operators on the subsystem on the left($\sigma^{\beta}$) to Majorana Basis $M_{L}^{\beta}$, which gives us the INSERT operator in Eq.(\ref{Insert}).
In regular recourse, we swap in the ancilla $0^{\mathrm{th}}$ qubit with the $1^{\mathrm{st}}$ qubit in our protocol. We will now study the impact of $\mathrm{SWAP(0,2)}$ on the final state and how it varies with $\mathrm{SWAP(0,1)}$. Let us denote $\mathrm{SWAP(0,1)}$ as $\Delta^{(0,1)}$ and $\mathrm{SWAP(0,2)}$ as $\Delta^{(0,2)}$ henceforth. The action of $\Delta^{(0,1)}$ is such that the $0^{\mathrm{th}}$ qubit, which was the message qubit, is swapped with the $1^{\mathrm{st}}$ qubit in the TFD state. In a similar fashion for $\Delta^{(0,2)}$, we have $0^{\mathrm{th}}$ qubit swapped with $2^{\mathrm{nd}}$ qubit. The fidelity expression Eq.(\ref{sqfide}) can now be stated as
\begin{equation}\label{eq:31}\begin{split}
     \expval{Z}_{R}^{\Delta^{(0,n_s)}} \propto Tr((Z_{R}e^{ig\hat{\upsilon}}e^{-iH_{L}t}\Delta^{(0,n_s)} \\ \ket{\psi}_{initial})(\bra{\psi}_{initial}\Delta^{(0,n_s)}e^{iH_{L}t}e^{-ig\hat{\upsilon}})) .
\end{split}
\end{equation}
These are generalized equations for different SWAP operators on the outcome of fidelity, wherein $n_s \in \{1,2\}$. They both differ in the positioning of the message qubit. The structure of the reduced density matrix is thus changed. We can see it more clearly if we further break down the generalized equations. Let us define an overlap coefficient by projecting the evolved initial state after $\Delta^{(0,n_s)}$ onto the TFD state basis and SIZE operator. We consider the overlap coefficient for the general case
\begin{equation}
\begin{split}
     C_{nm}^{(ij)}= \frac{1}{Z}\sum_{k1,k2}e^{-\beta (E_{k1}+E_{k2})/2}\bra{m_{L}}\gamma_{i}^{L}\ket{k_{1}} \\ \bra{m_{R}}\gamma_{j}^{R}\ket{k_{2}}\delta_{k_{1}k_{2}}.
\end{split}   
\end{equation}
The delta function ensures left and right indices remain correlated. For $\Delta^{(0,1)}$ case we get
\begin{equation} \label{cnm01}
    C_{nm}^{(01)} = \frac{1}{Z}\sum_{k}e^{-\beta E_{k}}\bra{m_{1}}\gamma_{1}^{L}\ket{k}\bra{m_{0}}\gamma_{1}^{R}\ket{k}.
\end{equation}
We see that Majorana bilinear term $\gamma_{1}^{L}\gamma_{1}^{R}$ dominates in the above equation. Similarly for $\Delta^{(0,2)}$ case,
\begin{equation} \label{cnm02}
     C_{nm}^{(02)} = \frac{1}{Z}\sum_{k}e^{-\beta E_{k}}\bra{m_{2}}\gamma_{2}^{L}\ket{k}\bra{m_{0}}\gamma_{2}^{R}\ket{k}.
\end{equation}
Now, the way Majorana operators are arranged in the SYK model, there is a specific distance dependence as these Majorana operators are paired into qubits \cite{chew2017approximating}. These qubits corresponding to Majorana operators with smaller indices are spatially closer to the message qubit. Correlation strength between message qubit operator and swapped qubit operator decreases with increasing distance in operator space \cite{gu2017local, huang2019eigenstate,lantagne2020diagnosing}
\begin{equation}
    Corr(n_{0},n_{k}) \propto\expval{\gamma_{i}^{L}\gamma_{j}^{R}} \equiv e^{-\lambda d_{ij}} .
\end{equation}
Here, $d_{ij}$ is the distance between qubits in their Majorana mode indices, and $\lambda$ is a decay constant. This decay follows exponential suppression. It follows that the fundamental correlation function between two Majorana operators in the SYK model is given as
\begin{equation}
    \expval{\gamma_{i}^{L}\gamma_{j}^{R}}_{\beta} = \frac{1}{Z}\sum_{k}e^{-\beta E_{k}}\bra{k}\gamma_{i}^{L}\gamma_{j}^{R}\ket{k} .
\end{equation} 
The SYK Hamiltonian contains all-to-all random couplings of Majorana fermions, which introduce disorder-averaged correlations \cite{maldacena2016remarks}. We define a relative phase factor $\phi$ that depends on interaction terms and, by extension, the distance between the Majorana fermion operators.
\begin{equation}
    \expval{\gamma_{i}^{L}\gamma_{j}^{R}}_{\beta} \sim e^{-\lambda d_{ij}} e^{i\phi_{k}} .
\end{equation}
These microscopic phases $\phi_k$ arise due to eigenstate interference, which is a product of the chaotic SYK spectrum. It is fairly straightforward to see that 
\begin{equation}
    C_{nm}^{(ij)} \propto e^{-\lambda d_{ij}} e^{i\phi_{nm}^{(ij)}} .
\end{equation}
Here, $\phi_{nm}^{(ij)}$ are effective phases arising from sum over $k$ eigenstates from Eq.~(\ref{cnm01}) and Eq.~(\ref{cnm02}). These effective phases govern the interference pattern that determines the teleportation fidelity. The fidelity is proportional to the overlap coefficient, given as
\begin{equation}
    \expval{Z}_{R}^{(ij)} \propto \sum_{n,m}e^{-\lambda d_{ij}} e^{i\phi_{nm}^{(ij)}} Z_{nm}^{R} ,
\end{equation}
where $Z_{R}^{mn}= \bra{m}Z_{R}\ket{n}$ is the expectation value of the Z operator in the right subsystem for $\ket{m}$ and $\ket{n}$ basis states.  \\
We now analyze the Random phase approximation for distance decay correlations of $\Delta^{(i,j)}$. From definitions of $C_{nm}^{(ij)}$ we define wavefunction coefficients and expand them as

\begin{equation}
\begin{split}
    \bra{m_{L}}\gamma_{i}^{L}\ket{k}= c_{m_{L}k}e^{i\phi_{ik}} , \\
     \bra{m_{R}}\gamma_{j}^{R}\ket{k}= c_{m_{R}k}e^{i\phi_{jk}}.
\end{split}
\end{equation}
We have $c_{m_{L}k}$, $c_{m_{R}k}$ as real wavefunction coefficients and $\phi_{ik}$,$\phi_{jk}$ as random phases induced by chaotic nature of SYK dynamics. Now, we expand the overlap coefficient as
\begin{equation}
    C_{nm}^{(ij)} = \sum_{k} e^{-\beta E_{k}} c_{m_{L}k} c_{m_{R}k}e^{i(\phi_{ik}+\phi_{jk})}.
\end{equation}
We now evaluate the overlap coefficient $C_{nm}^{(01)}$ for $\Delta^{(0,1)}$ case at $\beta=0$. For this case, Majorana operators are relatively close to each other owing to $e^{-\lambda}$ where $d_{ij} \xrightarrow{}1$. Using the Random Phase approximation, we can imply that the phase fluctuations are correlated for small distances in SYK model dynamics \cite{sonner2017eigenstate}. This means $\phi_{0k} \thickapprox \phi_{1k}$ so the phase term simplifies to
\begin{equation} \label{corr1}
    e^{i(\phi_{0k}+\phi_{1k})} = e^{2i\phi_{0k}} ,
\end{equation}
As phase fluctuations remain correlated, the sum does not cause destructive interference, which results in $\Delta^{(0,1)}$ finite correlation at $\beta=0$
\begin{equation}
    C_{nm, \beta = 0}^{(01)} = \sum_{k}c_{m_{L}k} c_{m_{R}k}e^{2i\phi_{0k}}.
\end{equation}
However, for $\Delta^{(0,2)}$, the distance between the two Majorana operators increases as the target qubit is relatively far from the original site in the system. There is a one-to-one correlation of quantum chaos and RMT \cite{bohigas1984characterization}, so due to the chaotic nature of the SYK model, its eigenstates resemble those of a random matrix ensemble \cite{erdHos2012spectral,garcia2022symmetry}. Here, the eigenstate coefficients behave like Gaussian-distributed random variables with zero mean. Their phase is also distributed randomly and no longer correlated for the two Majorana operators \cite{maldacena2016remarks}. The random phase approximation now tells us that phase fluctuations decohere for large distances $(\phi_{0k} \neq \phi_{2k})$. As phases are randomly distributed, $e^{i(\phi_{0k}+\phi_{2k})}$ randomly oscillates between $\pm{1}$. This causes the summed overall phases to average out to zero as random phases cancel each other \cite{gu2017spread}
\begin{equation} \label{corr2}
    \sum_{k}e^{i(\phi_{0k}+\phi_{2k})} \thickapprox 0 .
\end{equation}
Thus, we get destructive interference, so the overlap coefficients vanish to a null value.
\begin{equation}
    C_{nm, \beta = 0}^{(02)} = \sum_{k}c_{m_{L}k} c_{m_{R}k}e^{i(\phi_{0k}+\phi_{2k})} \thickapprox0 .
\end{equation}
As we go into the intermediate $\beta$ regime, the sum is weighted by the Boltzmann factor $e^{-\beta E_{k}}$, which suppresses high energy states' contribution to chaotic interference for $\Delta^{(0,2)}$. This reduces phase randomness because the strongly weighted terms have lower energies. These terms have correlated phases, resulting in partial constructive interference leading to
\begin{equation}
    \sum_{k} e^{i(\phi_{0k}+\phi_{2k})}e^{-\beta E_{k}} \neq0.
\end{equation}
Mathematically we approximate overlap coefficient for $\Delta^{(0,2)}$ case 
\begin{equation}
    C_{nm,\beta>0}^{(02)} \thickapprox \sum_{k} e^{i(\phi_{0k}+\phi_{2k})}e^{-\beta E_{k}} 
\end{equation}

We will now analyze how varying the $\beta$ regime quantitatively impacts the teleportation fidelity.

\subsubsection{Behavior at $\beta = 0$}
\begin{figure}[H]
    \centering
    \includegraphics[width=1.0\linewidth]{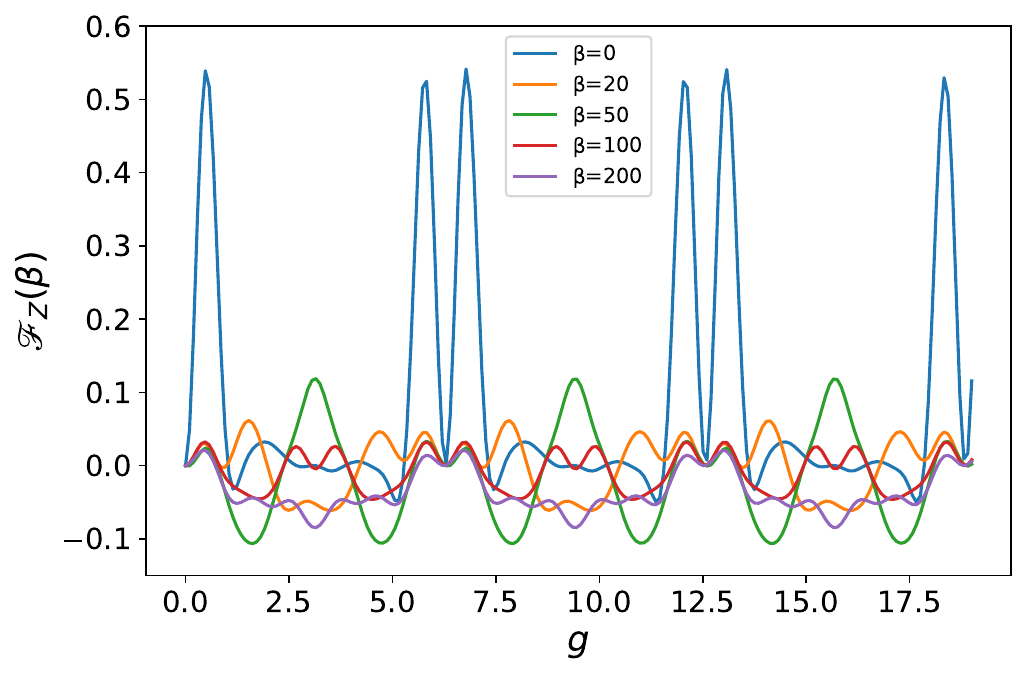}
    \caption{Teleportation fidelity for N=6 SYK Hamiltonian with $\Delta^{(0,1)}$ operator for a range of $\beta$ values. Fidelity is maximum for $\beta=0$ and reduces considerably for increasing values of $\beta$.}
    \label{sq1}
\end{figure}

\begin{figure}[H]
    \centering
    \includegraphics[width=1.0\linewidth]{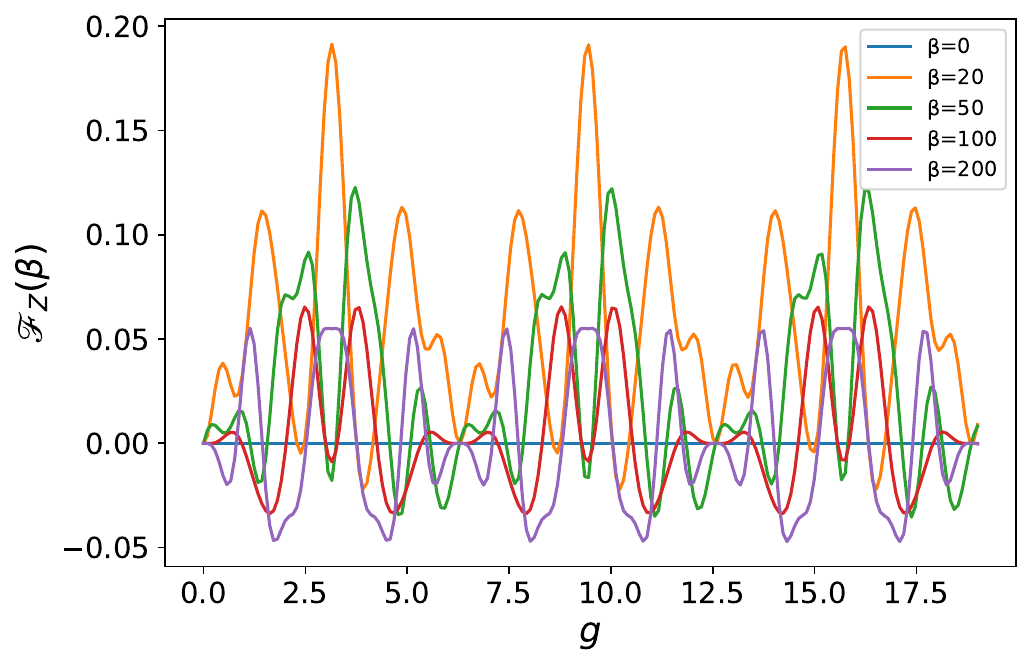}
    \caption{Teleportation fidelity for N=6 SYK Hamitonian with $\Delta^{(0,2)}$ operator for a range of $\beta$ values. Here, fidelity is minimum for $\beta=0$ and peaks at $\beta=20$, decreasing thereafter.}
    \label{sq2}
\end{figure}

\begin{figure}[H]
    \centering
    \includegraphics[width=1.0\linewidth]{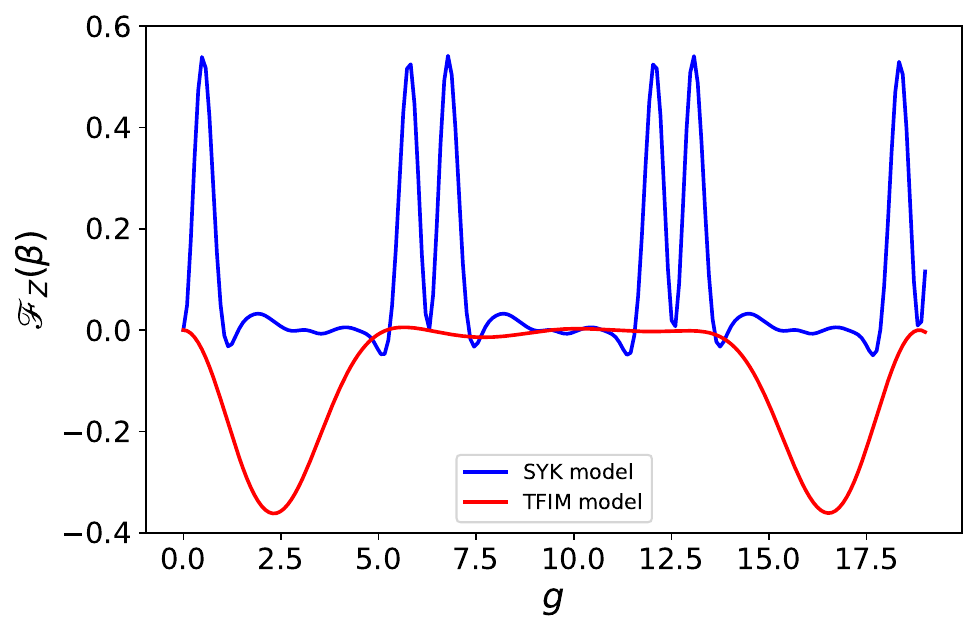}
    \caption{Comparison of teleportation fidelity for N=6 SYK Hamiltonian based Model and N=6 TFIM Hamiltonian Based model sweeping over values of $g$ at $\beta = 0$.}
    \label{isingvssyk}
\end{figure}

At $\beta=0$, the thermal state becomes a maximally mixed state, and the uniform superposition of states ensures that overlaps between swapped states and TFD basis states are non-zero due to the symmetry of the TFD state for $\Delta^{(0,1)}$. This regime has significantly strong correlations; hence, it effectively gives the maximum fidelity, as shown in Fig.(\ref{sq1}). However, for $\Delta^{(0,2)}$, the operation exchanges the $0^{\mathrm{th}}$ and $2^{\mathrm{nd}}$ qubits in the same subsystem, breaking the left-right symmetry, thus causing misalignment in TFD overlap. The SWAP operator destroys the correlations between the left and right subsystems inherent in the TFD state, which ceases information propagation \cite{shenker2014black}. Hence, $C_{nm}^{(02)}=0$ which leads to $\mathcal{F}_z(\beta)=\expval{Z}_{R}^{\Delta^{(0,2)}} =0 $ as demonstrated in Fig.(\ref{sq2}). We can also see that the teleportation fidelity of the TFIM-based model is substantially lower compared to the SYK counterpart shown in Fig.(\ref{isingvssyk}). TFIM-based model used is a self-dual Floquet Ising model with $J=b=\pi/4$ and $h_i$ is drawn from a Gaussian distribution of width $0.5$. Concretely, the Floquet operator is defined as \cite{bertini2019entanglement,brown2019quantum}:
\begin{equation}
    U_{\rm TFIM} = exp(ib\sum_iX_i)exp(iJ\sum_iZ_iZ_{i+1} + ih_i\sum_iZ_i)
\end{equation}
The reason for the lower teleportation fidelity of the TFIM-based protocol compared to the SYK model-based protocol is that TFIM dynamics are less chaotic and less ergodic, resulting in weaker scrambling and reduced effective left-right coupling.

\subsubsection{Behavior at intermediate $\beta > 0$ regime }
As the inverse temperature value increases, the TFD state's coherence undergoes dilution \cite{chapman2019complexity}. The thermal correlations thus diminish with an increase in $\beta$, leading to reduced entanglement \cite{su2021variational}. Also, the high-energy states are suppressed due to the Boltzmann factor acting as a weight and prioritizing low-energy states that are less affected by the SWAP operation. However, since $C_{nm}^{(01)}\neq 0$ even at $\beta=0$, the fidelity remains finite but reduced. The overlap $C_{nm}^{(02)}$ becomes non-zero due to less pronounced destructive interference from $\Delta^{(0,2)}$. The SWAP $\Delta^{(0,2)}$ also moves the message qubit further in the system, introducing a distance-dependent decay in the overlap between the L and R subsystems. At intermediate $\beta$ value, thermal correlations are maximized specifically for $\Delta^{(0,2)}$  This effectively leads to maximized teleportation fidelity for $\Delta^{(0,2)}$.

\subsubsection{Behavior at large $\beta>>1$ regime}
For $\Delta^{(0,1)}$ and $\Delta^{(0,2)}$, the thermal correlations decrease for both cases as the ground state dominates the system. The overlap between two subsystems approaches a near-separable ground state, resulting in non-trivial but low teleportation fidelity.

\section{Temporal fluctuation of fidelity in the Bulk of AdS/CFT boundary}
We define a message recovery time $t_{\mathrm{recovery}}$ as when the teleported qubit state on the right side matches most closely with the original message qubit inserted on the left side \cite{hayden2007black}. This can be mathematically tied to internal dynamics of the expectation value of $Z_{R}$ operator and fidelity of the final state \cite{yoshida2017efficient}. 
Spectral gap ($\Delta_{s}$) \cite{tezuka2023binary} plays a crucial role in determining $t_{\mathrm{recovery}}$. Here, we derive how the spectral gap of the $\hat{\upsilon}$ operator and the coupling affect the recovery time. From Eq.(\ref{size}) let $\ket{s}$ denote an eigenstate of $\hat{\upsilon}$ with eigenvalue $s$
\begin{equation}
    \hat{\upsilon}\ket{s}=s\ket{s}.
\end{equation}
We have effectively diagonalized in the eigen basis of $\hat{\upsilon}$. The spectrum of the SIZE operator determines how the system responds to the coupling term $e^{ig\hat{\upsilon}}$. The eigenvalues $s$ represent the quantized Size of the system for a given eigenstate. The eigenstates $\ket{s}$ form a complete basis for Hilbert space, and they encode the structure of the $\hat{\upsilon}$ operator, which directly relates to the wormhole-inspired teleportation protocol by measuring how much information is localized in the left compared to the right subsystem. The spectral gap $\Delta_{s}$ is the difference between the smallest and second smallest non-zero eigenvalue of the SIZE operator \cite{polchinski2016spectrum}.
   $ \Delta_{s}=s_{2}-s_{1}$.
The term $e^{ig\hat{\upsilon}}$ causes a phase shift in eigenstates of $\hat{\upsilon}$
\begin{equation}
    e^{ig\hat{\upsilon}}\ket{s} = e^{igs}\ket{s}.
\end{equation}
For a smaller $g$ value, this induces slow dynamics between eigenstates of the SIZE operator. The recovery time $t_{\mathrm{recovery}}$ depends on how quickly the system evolves into the state where the qubit's information is transferred from the left to the right subsystem. A spectral gap determines this rate. Now, the evolution operator can be written in the eigenbasis of $\hat{\upsilon}$ as 
\begin{equation}
    U_{\mathrm{wormhole}}(t)= \sum_{s} e^{-i(E_{R}+gs)t}\ket{s}\bra{s},
\end{equation}
 where $E_{R}$ is the contribution from right subsystem Hamiltonian $H_{R}$. Our calculation of teleportation fidelity in the previous sections involved fixing the time and varying the coupling strength. Now we proceed to fix the coupling strength at the maximum value of $g$ for a particular $\beta$ case and calculate the fidelity as a function of time ($t$) as
    
\begin{equation}
    \expval{Z(t)}_{R} = \bra{\psi_{\mathrm{final}}(t)}Z\ket{\psi_{\mathrm{final}}(t)} .
\end{equation}
where $\ket{\psi_{\mathrm{final}}(t)} = U_{\mathrm{wormhole}}(t)\ket{\psi_\mathrm{initial}} $. The message recovery time $t_{\mathrm{recovery}}$ depends on how fast the phases ($\phi=gst$) collect for different eigenvalues $s$. The interference between states with consecutive eigenvalues $s_{2}$ and $s_{1}$ also contributes to the recovery process. Mathematically, the phase difference is 
\begin{equation}
    \delta=g(s_{2}-s_{1})t=g\Delta_{s}t .
\end{equation}
Constructive interference occurs at maximum recovery time such that
\begin{equation}
    g\Delta_{s}t_{\mathrm{recovery}} \sim \pi. 
\end{equation}
In terms of our protocol conditions, recovery time corresponds to 
\begin{equation}
    t_{\mathrm{recovery}} = \underset{t}{\mathrm{argmax}}\expval{Z(t)}_{R} .
\end{equation}
The peak occurs as the message qubit's information fully transfers from the left to the right subsystem. We will now try to find how exactly the recovery time varies with variational $\beta$ for $\Delta^{(0,1)}$ and $\Delta^{(0,2)}$ cases. In our protocol setup, Majorana modes on left and right subsystems are organized in such a way that their correlations decrease with spatial separation \cite{sonner2017eigenstate},\cite{gu2017spread}
\begin{equation}
    \expval{\gamma_{L}^{(i)}\gamma_{R}^{(j)}} \sim e^{-\lambda d_{ij}}.
\end{equation}
For $\Delta^{(0,1)}$, the distance between the left and right Majorana modes is smaller, leading to strong correlations compared to $\Delta^{(0,2)}$. This can be shown by considering the diagonal correlation as
\begin{equation}
    C_{nm}^{(ij)}=\bra{TFD}\gamma_{L}^{(i)}\gamma_{R}^{(j)}\ket{TFD}.
\end{equation}
It is rather easy to see one-to-one correspondence between the $\Delta_{s}$ and $C_{ij}$ as a change in Size is akin to a correlation function between corresponding Majorana modes.
\begin{equation}
    t_{\mathrm{recovery}}^{(0,n_{s})} \propto \frac{\pi}{g \bra{TFD}m_{L}^{(0)}m_{R}^{(n_{s})}\ket{TFD}}.
\end{equation}
Since the correlation function for $\Delta^{(0,2)}$ is smaller due to greater physical separation compared to $\Delta^{(0,1)}$ mode indices from Eq.~(\ref{corr1}) and Eq.~(\ref{corr2}), we have
\begin{equation} \label{rec}
    t_{\mathrm{recovery}}^{(0,2)} > t_{\mathrm{recovery}}^{(0,1)}.
\end{equation}

\begin{figure} 
    \centering
    \includegraphics[width=1.0\linewidth]{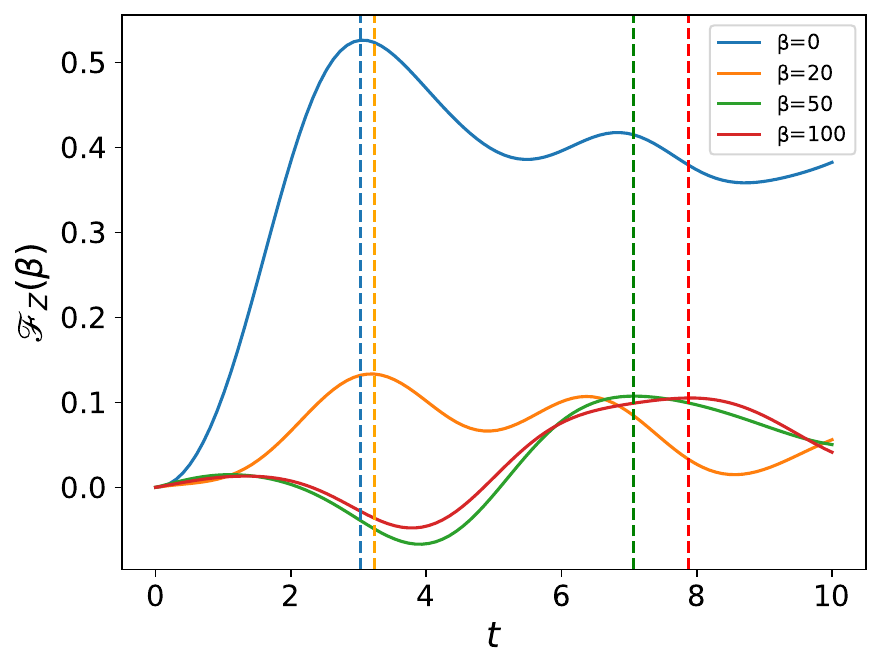}
    \caption{Variation of teleportation fidelity for N=6 SYK Hamiltonian and $\Delta^{(0,1)}$ operator with time.}
    \label{fig:6}
\end{figure}

\begin{figure}
    \centering
    \includegraphics[width=1.0\linewidth]{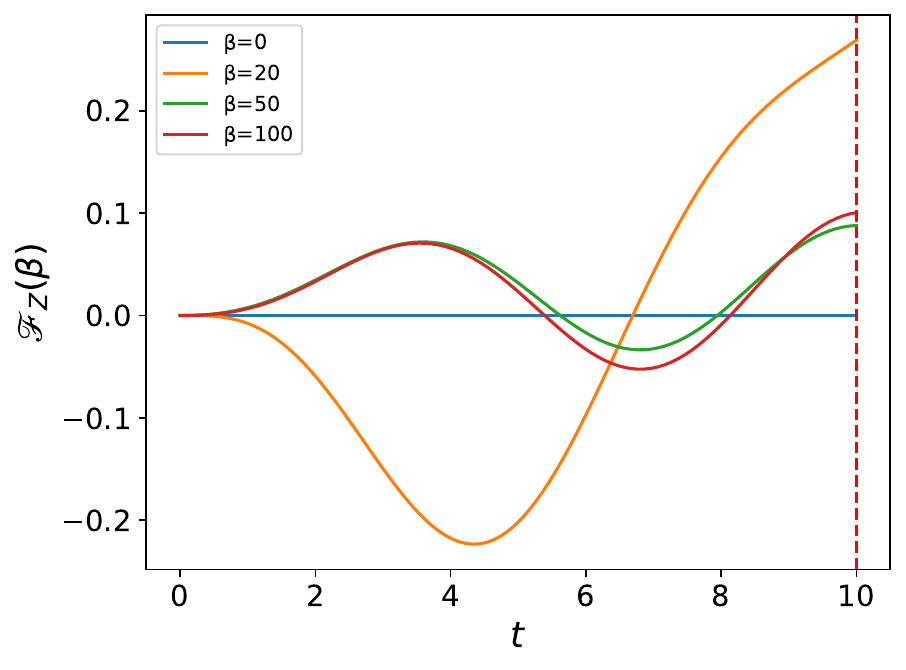}
    \caption{Variation of teleportation fidelity for N=6 SYK Hamiltonian and $\Delta^{(0,2)}$ operator with time. For $\beta=0$, fidelity is ``flat-lined".}
    \label{fig:7}
\end{figure}

Data from Figs.(\ref{fig:6}), and (\ref{fig:7}) make it clear that the fidelity variation for both the SWAP operator cases conforms to Eq.(\ref{rec}).

\section{Bell state teleportation in Ads/CFT analogous wormhole-inspired teleportation protocol} \label{bell}
For two-qubit teleportation, we use the Bell state $\ket{\Phi}^+ $ as our message state and then evolve it unitarily to achieve teleportation and ascertain the protocol's fidelity. The form of the rest of the protocol sections is akin to that of a single-qubit teleportation case, except for the SWAP operator structure.
In the case where the message is the Bell state $|\Phi^+\rangle = \frac{1}{\sqrt{2}}(|00\rangle + |11\rangle)$, the teleportation fidelity is evaluated via a Pauli stabilizer-based approach \cite{bussandri2024challenges, kalev2019validating, flammia2011direct}. For this, we define a new Pauli stabilizer fidelity:
\begin{equation} \label{fidestab}
\mathcal{F}_{\Phi_{+}}(\beta) = \frac{1}{2} \left(1 + \langle S_1 \rangle + \langle S_2 \rangle + \langle S_3 \rangle \right),
\end{equation}
where $S_1 = X \otimes X$, $S_2 = Z \otimes Z$, and $S_3 = Y \otimes Y$ are the stabilizers of the $|\Phi^+\rangle$ state. For testing the robustness of this new fidelity approach specifically for the Bell triplet state, if we were to find the fidelity for the initial state, i.e., just the initial state reduced density matrix, we would get a perfect fidelity (Refer to Appendix \ref{appendix1}). For actual fidelity calculations of the two-qubit protocol after unitary evolution through bulk and boundary Fig.(\ref{BSQC}), the expectation values are computed over the two-qubit reduced density matrix obtained from the final state of the protocol as 
\begin{equation}
    \expval{S_1} = \bra{\psi_{\mathrm{final}}}S_1\ket{\psi_{\mathrm{final}}}  .
\end{equation}
A similar calculation is performed for $\expval{S_2}$ and $\expval{S_3}$, and the fidelity is thus calculated for Bell state teleportation.

\begin{figure*} [t]
\centering
    \begin{tikzpicture} [scale=1.0]
\draw[very thick] (-4,0.5) -- (4,0.5);
\draw[very thick] (-4,0) -- (4,0);
\draw[very thick]  (-4,-0.5) -- (4,-0.5);
\draw[very thick] (-4,-1) -- (4,-1);
\draw[very thick]  (-4,-1.5) -- (4,-1.5);
\draw[very thick] (-4,-2) -- (4,-2);
\draw[very thick] (-4,-2.5) -- (4,-2.5);
\draw[very thick] (-4,-3) -- (4,-3)node[below right] {$\ket{\psi}_{final}$};

\node[draw=black, fill=white, rounded corners=5pt, minimum width=1cm, minimum height=1.5cm, align=center] at (-3,-1) {$e^{iH_{L}t}$};
\node[draw=black, fill=white, rounded corners=5pt, minimum width=1cm, minimum height=1.5cm, align=center] at (-0.625,-1) {$e^{-iH_{L}t}$};
\node[draw=black, fill=white, rounded corners=5pt, minimum width=1cm, minimum height=1.5cm, align=center] at (3,-2.5) {$e^{-iH_{R}t}$};
\node[draw=black, fill=white, rounded corners=5pt, minimum width=1cm, minimum height=2.25cm, align=center] at (1,-1.75) {$e^{igV}$};

\draw [decorate,decoration={brace,amplitude=10pt,mirror}] (-4,0.5)  -- (-4,0) node[midway,xshift=-25pt] { $\ket{\Phi^{+}}$} ;

  \node at (-2.25,0.5) {$\times$};
  \node at (-2.25,-0.5) {$\times$};
   \node at (-1.75,0) {$\times$};
  \node at (-1.75,-1) {$\times$};

  \draw (-2.25,0.6) -- (-2.25,-0.6);
   \draw (-1.75,0) -- (-1.75,-1);

  \node[above] at (-2.25,0.6) {\small SWAP(0,2)};
  \node[above] at (-1.4,0) {\small SWAP(1,3)};

 \draw[very thick] (-4,-0.5) arc[start angle=-270, end angle=-90, radius=1.25 cm];
 \draw[very thick] (-4,-1) arc[start angle=-270, end angle=-90, radius=0.75 cm];
  \draw[very thick] (-4,-1.5) arc[start angle=-270, end angle=-90, radius=0.25 cm];

  \node[fill=red, fill opacity=0.3, text opacity=1, draw=blue, rounded corners, minimum width=2cm, minimum height=1.5cm, align=center] at (-5,-0.875)
    {\Huge L};

  \node[fill=red, fill opacity=0.3, text opacity=1, draw=blue, rounded corners, minimum width=2cm, minimum height=1.5cm, align=center] at (-5,-2.625)
    {\Huge R};

\end{tikzpicture}
\caption{Quantum circuit representation of Bell state Wormhole-Inspired teleportation protocol. Here, $\ket{\psi_{\rm initial}}= \ket{\Phi^{+}} \otimes \ket{TFD}$ and the rest of the protocol follows the same notation as single qubit case with difference being the successive SWAP operations that push the message into the bulk.}
 \label{BSQC}
\end{figure*}

To understand the $\beta$-dependence, we define the thermal overlap coefficients
\begin{equation}
C_{nm}^{\text{Bell}}(\beta) = \sum_k e^{-\beta E_k} \langle \psi_k | P_n^\dagger | \psi_k \rangle \langle \psi_k | P_m | \psi_k \rangle,
\end{equation}
where $\ket{\psi_k}$ are the eigenstates of the SYK Hamiltonian and $P_n, P_m$ are two-qubit Pauli operators \cite{bagarello2024pauli} appearing in the decomposition of the decoding SWAP operator. We have defined the SWAP operator here as a sequential SWAP to account for both qubit message channels, which is decomposed as $\mathrm{SWAP}= \sum_{n} c_nP_n$ or equivalently, $\mathrm{SWAP}=\sum_{m,n}c_{mn}P_{n}^{\dagger}P_{m}$ and notably the coefficients in expanded Pauli basis are $c_n=\frac{1}{4}Tr(P_{n}^{\dagger}\mathrm{SWAP})$. This coefficient is like a weight representing how much the Pauli string $P_n$ contributes to the SWAP operator on a Pauli-expanded basis.

\begin{figure}[H] 
    \centering
    \includegraphics[width=1.0\linewidth]{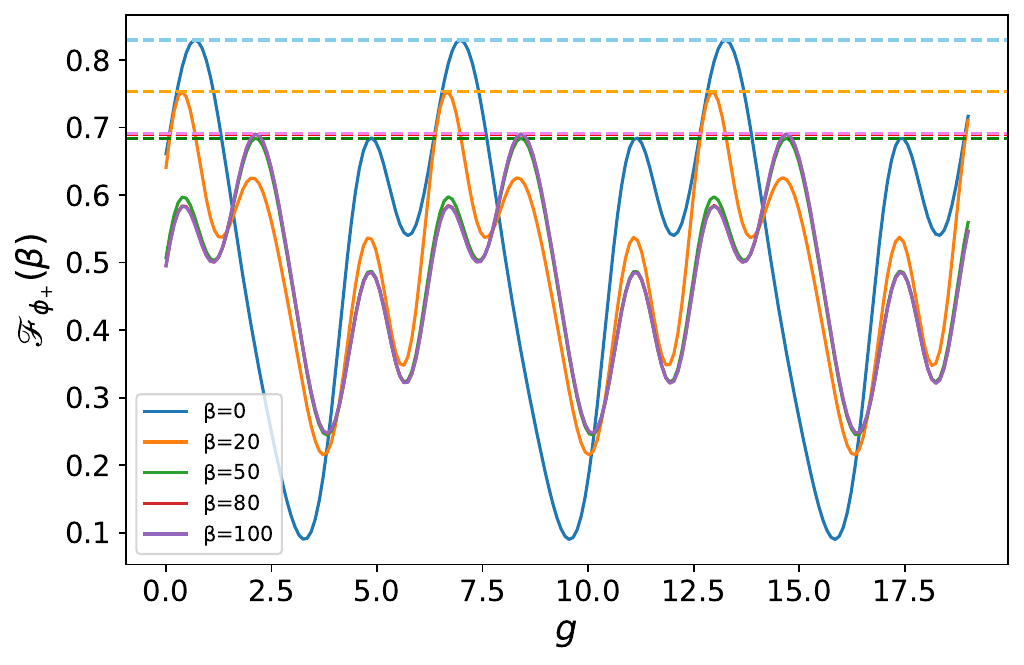}
    \caption{Variation of teleportation fidelity for Bell state as a function of coupling strength ($g$).}
    \label{fig34}
\end{figure}

\begin{figure}[H]
    \centering
    \includegraphics[width=1.0\linewidth]{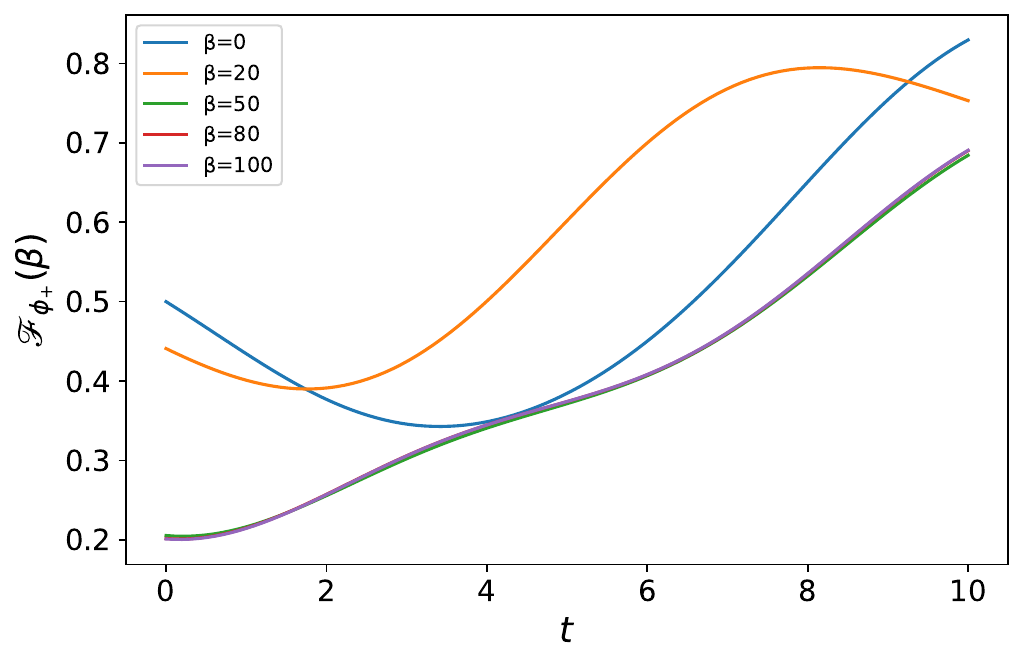}
    \caption{Variation of teleportation fidelity for Bell state as a function of traversal time ($t$).}
    \label{timeevol}
\end{figure}

Now, in order to find an expression for fidelity based on overlap coefficients, we define a teleportation channel $ \varepsilon$ such that
\begin{equation}
    \varepsilon(\rho) = \sum_{n,m}c_{nm}(\beta) P_{n}\rho P_{m}^{\dagger} ,
\end{equation}
Post teleportation, we have $\rho_{teleported}=\varepsilon(\rho)$ and where $\rho = \ket{\Phi^{+}}\bra{\Phi^{+}}$ then,
\begin{equation}
    \rho_{\mathrm{teleported}} = \sum_{n,m}c_{nm}(\beta)P_{n}\ket{\Phi^{+}}\bra{\Phi^{+}}P_{m}^{\dagger} .
\end{equation}
For each Pauli stabilizer $S_{i} \in \{X \otimes X, Y \otimes Y, Z \otimes Z\}$ we have $\expval{S_i}$ which is
\begin{equation}
     Tr(\rho_{\mathrm{teleported}} S _i) = \sum_{n,m}c_{nm}(\beta) Tr(P_n\ket{\Phi^{+}}\bra{\Phi^{+}}P_{m}^{\dagger} S_i) .
\end{equation}
Using the cyclic property of trace and using the fact that
   $  \bra{\Phi^{+}} P_{m}^{\dagger} S_iP_n\ket{\Phi^{+}} = Tr(S_iP_n^{\dagger}P_m) /4$,
we write the expression for Pauli stabilizers
\begin{equation}
    \expval{S_i} = \sum_{n,m}c_{nm}(\beta) Tr(S_iP_n^{\dagger}P_m).
\end{equation}
The above equation connects fidelity to SYK dynamics via overlap coefficients and Bell state stabilizer-based structure.
The fidelity decays with increasing $\beta$ despite fixed physical qubit positions due to thermal phase decoherence induced by the SYK Hamiltonian \cite{kobrin2021many}, as shown in Fig.(\ref{fig34}). Thus, we can see that fidelity is sensitive to how well the thermalized channel preserves the coherence of the Bell message. From Fig.(\ref{timeevol}), we can see that fidelity maxes out for almost all $\beta$ values towards the end of the protocol. This reflects a coherent buildup of teleportation amplitude in the presence of SYK chaotic dynamics and thermal suppression.

\begin{figure}[H]
    \centering
    \includegraphics[width=1.0\linewidth]{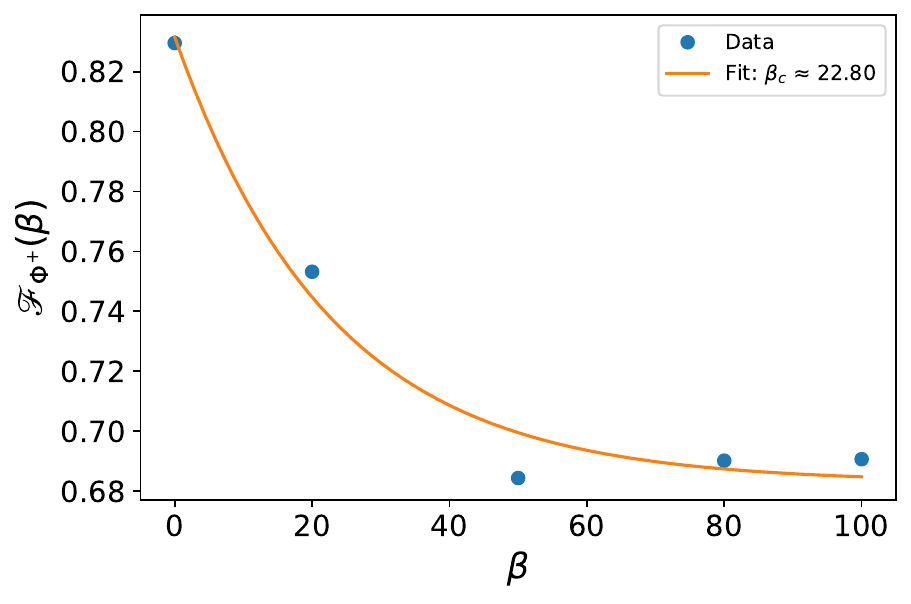}
    \caption{Effective thermal coherence suppression scale plotted as a curve fit in a range of $\beta$ values and $\mathcal{F}_{\Phi^{+}}$ for Bell state teleportation protocol.}
    \label{cffit}
\end{figure}

\begin{figure}[H]
    \centering
    \includegraphics[width=1.0\linewidth]{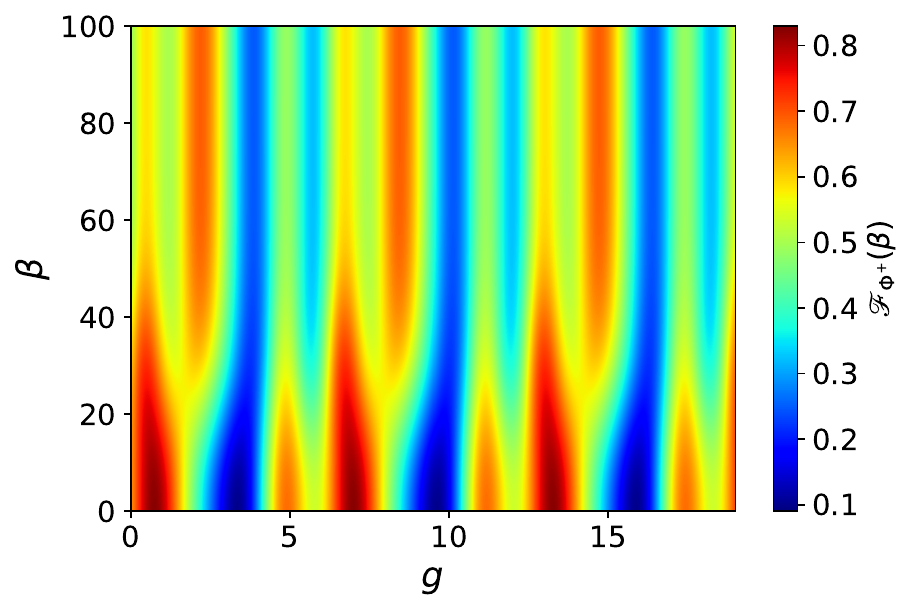}
    \caption{Heatmap of Bell state teleportation fidelity as a function of $\beta$ values and  coupling strength ($g$).}
    \label{2dfideg}
\end{figure}

\begin{figure}[H]
    \centering
    \includegraphics[width=1.0\linewidth]{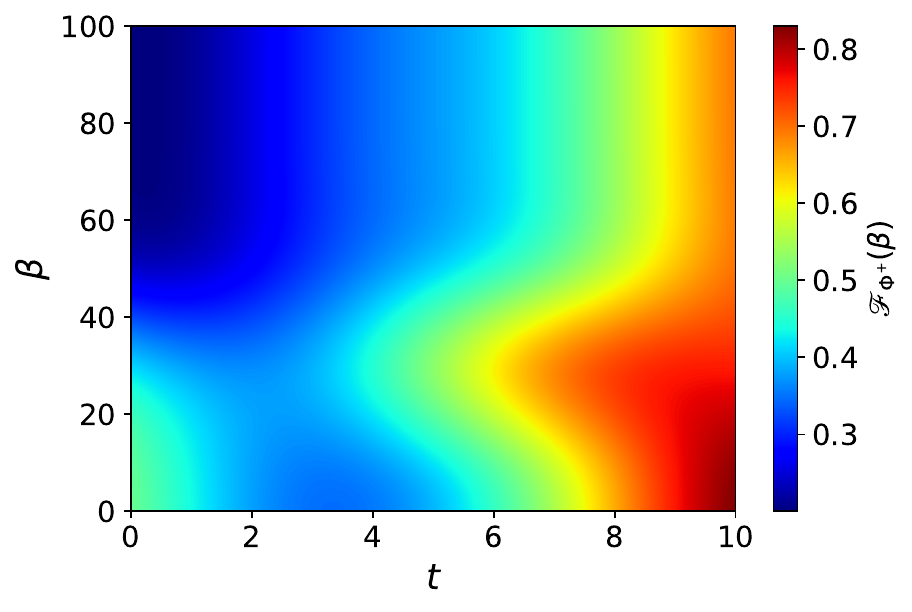}
    \caption{Heatmap of Bell state teleportation fidelity as a function of $\beta$ values and traversal time ($t$).}
    \label{2dfidet}
\end{figure}

Due to the chaotic nature of the SYK model, the eigenstate matrix elements acquire rapidly fluctuating complex phases $\phi_k^{(n,m)}$ \cite{chowdhury2022sachdev}. At $\beta = 0$, contributions from all energy levels interfere constructively on average due to random phase averaging \cite{sonner2017eigenstate}
leading to a maximum fidelity at infinite temperature. As $\beta$ increases, thermal suppression restricts the sum to a narrower band of eigenstates, but the chaotic phase structure persists, leading to destructive interference and:
\begin{equation}
C_{nm}^{\text{Bell}}(\beta) \sim e^{-\beta / \beta_c},
\end{equation}
for some effective coherence scale $\beta_c$ \cite{gu2017local}. In the SYK model, usually $\beta_c \sim \frac{1}{J}$ where J is the interaction strength drawn from a Gaussian random distribution as seen in Eq.(\ref{HL}) and Eq.(\ref{HR}). As we can see computationally from Fig. (\ref{fig34}), the fidelity decreases exponentially with $\beta$ and plateaus around $\beta \approx 80$. A visually clearer representation can be glanced from Fig.~(\ref{2dfideg}) and Fig.~(\ref{2dfidet}) where we can see the behaviour of teleportation fidelity across all the values of $\beta$,    $g$, and $t$ in their respective predefined range. Extrapolating from these figures, it is clear that the teleportation window occurs at periodically oscillating $g$ values and late times corresponding to specific $\beta$. Since $\beta_c$ determines how quickly teleportation fidelity decays with temperature \cite{jensen2016chaos}, we will extract it empirically by a simple curve fit. In the thermal SYK setup, thermal suppression of high-energy states is governed by Boltzmann weights \cite{bhattacharya2019quantum}. The Overlap coefficient and fidelity are dominated by low-energy states for large $\beta$. This transition happens around a characteristic decay scale $\beta_c$ where the fidelity decay slows down and saturates. Since we have already observed that $\mathcal{F}_{\Phi^{+}}$ decays exponentially with $\beta$ and saturates at large $\beta$, we will approximate this using a simple non-linear regression curve-fit by
\begin{equation}
    \mathcal{F}_{\Phi^{+}} (\beta) \approx A + Be^{-\frac{\beta}{\beta_c}} .
\end{equation}
Here, A is the asymptotic fidelity in the limit of $\beta\rightarrow\infty$, and B is the amplitude of initial thermal decay, which we denote as $ B=\mathcal {F}_{\Phi^{+}}(0) - A$. From Fig.(\ref{cffit}), we find $\beta_c \approx 22.80$ for this particular protocol. Interpreting it physically, we conclude that for $\beta << \beta_c$, the system is in a noiseless regime, and the teleportation fidelity is high because constructive interference is enhanced as there are many terms in the thermal sum \cite{failde2025hamiltonian}. Despite phases fluctuating randomly, correlation survives. When $\beta >> \beta_c$, the thermal sum peaks sharply near the ground state. However, due to the chaotic phase structure of SYK dynamics, even those low-energy states exhibit random phases, leading to destructive interference for random phases.\\
To place our finite $N$ numerical results in the context of established analytical work \cite{gao2017traversable, jafferis2022traversable,gao2021traversable}, we compare them with low-temperature, semiclassical analyses of wormhole teleportation in the SYK system. A direct quantitative comparison, such as plotting analytical curves on our figures, is not straightforward, as the key analytical predictions  are derived in the semiclassical limit where both the number of fermions $N$ and the interaction order $q$ are large. Our simulation, performed at finite $N=6$ and $q=4$, explores a physically distinct regime. Nevertheless, our findings are in strong qualitative agreement with the analytical expectations.\cite{gao2021traversable} showed that, in the low temperature/large-$N$ gravitational (semiclassical) regime limit of the coupled SYK-model, appropriately chosen left-right couplings and decoding operations can produce high mutual information $I(L:R)$: the analytical saddle supporting a traversable wormhole gives coherent size-winding cancellation and hence an $O(1)$ transmission amplitude. Subsequent studies that analyze operator spreading and chord-diagram semiclassical limits have confirmed that low temperature and semiclassical large-$N$ limit are the regimes in which fidelity peaks are sharp \cite{jafferis2022traversable}. Our finite-$N$, finite-$q$ ($N=6,q=4$) simulation displays the same qualitative tendency, which is that the fidelity increases as thermal suppression is reduced and a coherent size-winding condition is satisfied, producing fidelity peaks at specific coupling $g$ and $t$.


\section{Conclusion and Discussions}
In this work, we have implemented a sophisticated wormhole-inspired teleportation protocol based on SYK dynamics and TFD states. We began our exploration with single-qubit teleportation, thoroughly analyzing the influence of temperature on the protocol's fidelity. Our findings reveal a significant correlation between the physical distance between qubits and the outcomes of the teleportation process. We elucidated the dependence of temperature, physical distance, and teleportation fidelity by applying random phase approximations (RPA) derived from chaotic SYK model dynamics. We also compared teleportation fidelity for the TFIM-based and SYK-based models and found a higher fidelity for the latter. \\

Next, we progressed to a two-qubit protocol, embedding a $\Phi^+$ Bell state within the message qubits. We redefined the INSERT operator to accommodate an additional message channel, successfully demonstrating its wormhole-inspired teleportation. To accurately assess the fidelity of this Bell state teleportation protocol, we developed a comprehensive Pauli stabilizer-based fidelity calculation approach, which aligned seamlessly with the parameters of the protocol.
We then conducted rigorous fidelity diagnostics, incorporating Pauli stabilizer-based fidelity and analyzing their variance with coupling strength($\mathbf{g}$) and traversal time($\mathbf{t}$). Our results confirmed the highest fidelity($\mathcal{F}_{\Phi_{+}} = 0.83 $)for infinite temperature($\beta = 0$) regime, aligning perfectly with previous findings. Furthermore, we demonstrated that fidelity consistently peaks at later times for all $\beta$ for the Bell state version of the protocol and for  $\Delta^{(0,2)}$ of the single qubit version. Additionally, we established an effective coherence scale, $\beta_c$, which acts as a crucial transition point. Below this threshold, the system tends toward maximal coherent states, while above it, effective decoherence occurs. In summary, this work demonstrates effective Bell state teleportation through a traversable wormhole governed by SYK dynamics. It highlights the vital role of temperature and Majorana operator distance in regulating the transfer of quantum information within holography-inspired systems.

Our results also have direct bearing on the ER=EPR conjecture. The thermofield-double (TFD) state that underlies our two-sided SYK channel realizes an entangled left–right state whose low-energy bulk dual is commonly interpreted as two boundaries connected by a wormhole.  Within this perspective, the double-side coupling $e^{i g\hat\upsilon}$ that renders the channel traversable is the boundary manifestation of a traversable deformation in the bulk, and successful recovery of a Bell pair on the right corresponds to coherent information transfer through the emergent geometry.  Consequently, the present Bell-state teleportation protocol furnishes a simulation testbed for aspects of ER=EPR: (i) the mapping from left–right entanglement to geometric connectivity is probed by the TFD preparation; (ii) traversability and information transfer are diagnosed operationally by the Bell-state fidelity and its dependence on coupling $g$, time $t$, and temperature $\beta$; and (iii) the size-winding and phase-alignment mechanism we observe is precisely the microscopic channel that underpins entanglement-to-geometry intuition.  We emphasize that these numerical tests probe model-dependent and finite-$N$ realizations of the conjecture, rather than proving a general equivalence. Specifically, finite-size effects, finite $q$, and disorder averaging limit extrapolation are considered, which ultimately lead to the strict semiclassical (large-$N$, low-$\beta$) regime.  Nonetheless, the qualitative agreement between fidelity peaks, 2D heatmap structure, and the expected traversability windows supports the interpretation of our protocol as a useful and physically transparent simulator for probing ER=EPR phenomenology in microscopic many-body systems. We underscore, however, that these are finite-$N$, finite-$q$ and model-dependent tests: they are useful for exploring the conjecture’s concrete consequences in many-body systems but do not constitute a general proof of ER=EPR.

\section{Acknowledgement}

We would like to acknowledge the support provided by the project ``Study of quantum chaos and multipartite entanglement using quantum circuits" sponsored by the Science and Engineering Research Board (SERB), Department of Science and Technology (DST), India, under the Core Research Grant CRG/2021/007095.

\section{Data Availability}
The data that supports the findings of this paper is openly available at \cite{joshi2025witp}.

\bibliographystyle{unsrtnat}
\bibliography{MBL}

\appendix
\begin{widetext}
\section{}
\label{appendix2}
For arbitrary state input, the full initial state of the system is:
\begin{equation}
    \ket{\psi}_{\rm initial} = (\alpha\ket{0}+\beta\ket{1}) \otimes\ket{TFD}.
\end{equation}
This system then evolves under the wormhole unitary operator defined in Eq.~(\ref{wunitary}). By the principle of superposition, the final state of the system is a linear combination of the final states that would result from the individual computational basis inputs:
\begin{equation}
    \ket{\psi}_{\rm final} = U_{\rm wormhole}\ket{\psi}_{\rm initial} = \alpha U_{\rm wormhole}(\ket{0}\otimes\ket{TFD}) + \beta U_{\rm wormhole} (\ket{1}\otimes\ket{TFD}).
\end{equation}
Let us denote the final state of the entire many-body system for the input $\ket{0}$ as $\ket{\psi_{\rm final}(0)}$ and for input $\ket{1}$ as $\ket{\psi_{\rm final}(1)}$. The quality of teleportation is assessed by examining the reduced density matrix of the message qubit at output, $\rho_{out}$, which is obtained by tracing out SYK degrees of freedom ${\rm Tr}_{SYK}(\ket{\psi}_{\rm final}\bra{\psi}_{\rm final})$ which expands to:
\begin{equation} \label{rhoout}
    \rho_{\rm out} = |\alpha|^{2}\rho_{\rm out}(0) + |\beta|^{2}\rho_{\rm out}(1) +\alpha\beta^{*} {\rm Tr}_{SYK}(\ket{\psi_{\rm final}(0)}\bra{\psi_{\rm final}(1)}) +\alpha^{*}\beta {\rm Tr}_{SYK}(\ket{\psi_{\rm final}(1)}\bra{\psi_{\rm final}(0)}).
\end{equation}
where $\rho_{\rm out}(0)$ and $\rho_{\rm out}(1)$ are reduced density matrices for individual basis-state inputs. This formalism clarifies the distinction between two aspects of information transfer. The first two terms in Eq.~(\ref{rhoout}), proportional to classical probabilities $|\alpha|^2$ and $|\beta|^2$, govern the successful transmission of basis states themselves. The metric used in Eq.~(\ref{sqfide}), is a direct probe of this component, as it measures the probability of successfully recovering the initial basis state $\ket{0}$. 
The next two terms, the off-diagonal coherence, are responsible for preserving the quantum phase relationship between basis states. \\
The primary physical signature of a traversable wormhole is the successful reversal of information scrambling, which allows a signal to be recovered at all. This effect is most directly and cleanly probed by diagonal terms, i.e, the probability of recovering the correct basis state. The off-diagonal terms, which depend on the overlap between two distinct and complex many-body final states, are highly susceptible to decoherence from a thermal, chaotic environment. As we have put forth in Sec.~(\ref{3b}), the chaotic nature of SYK dynamics ensures that the energy eigenstate coefficients have random phases. Any overlap between two distinct, complex many-body states evolved under such dynamics will involve a sum over a large number of terms with uncorrelated random phases, leading to destructive interference that averages overlap to zero. This phenomenon, where the chaotic system effectively acts as a thermal, decohering environment, is a hallmark of quantum chaos and is expected to ``classicalize" the channel by projecting the output onto a statistical mixture.  \\
The teleportation fidelity for a single, arbitrary pure input state is the ``overlap" between the input state and the actual(generally mixed) output density matrix $\rho_{out}$:
\begin{equation}
    F(\ket{\psi}_{in}) = \bra{\psi_{in}}\rho_{out}\ket{\psi_{in}}.
\end{equation}
The fidelity is this quantity integrated over the entire Bloch sphere of possible input states:
\begin{equation}
    \mathfrak{F}_{\alpha\beta} = \int_{Bloch}F(\ket{\psi}_{in})d\psi
\end{equation}
The integral represents the average success of the protocol for any arbitrary quantum state. The fidelity for any single, random state $\ket{\psi}_{in}$ depends on its specific location on the Bloch sphere, leading to the fluctuations in single-run simulations. This is because the fidelity $\mathfrak{F}_{\alpha\beta}$ explicitly depends on the random input state coefficients $(\alpha,\beta)$ of that specific run. To obtain a meaningful metric for the channel itself, we must average this fidelity over a large ensemble of random input states. We denote this averaged fidelity as:
\begin{equation}
    \overline{\mathfrak{F}}_{\alpha\beta} = \frac{\sum_{i=1}^{n_s}\mathfrak{F}_{\alpha\beta}}{n_s}.
\end{equation} 
Here, $n_s$ represents the total numerical sumulations or the total number of iterations over which the mean is calculated.
\begin{figure}[H]
    \centering
    \includegraphics[width=0.49\linewidth]{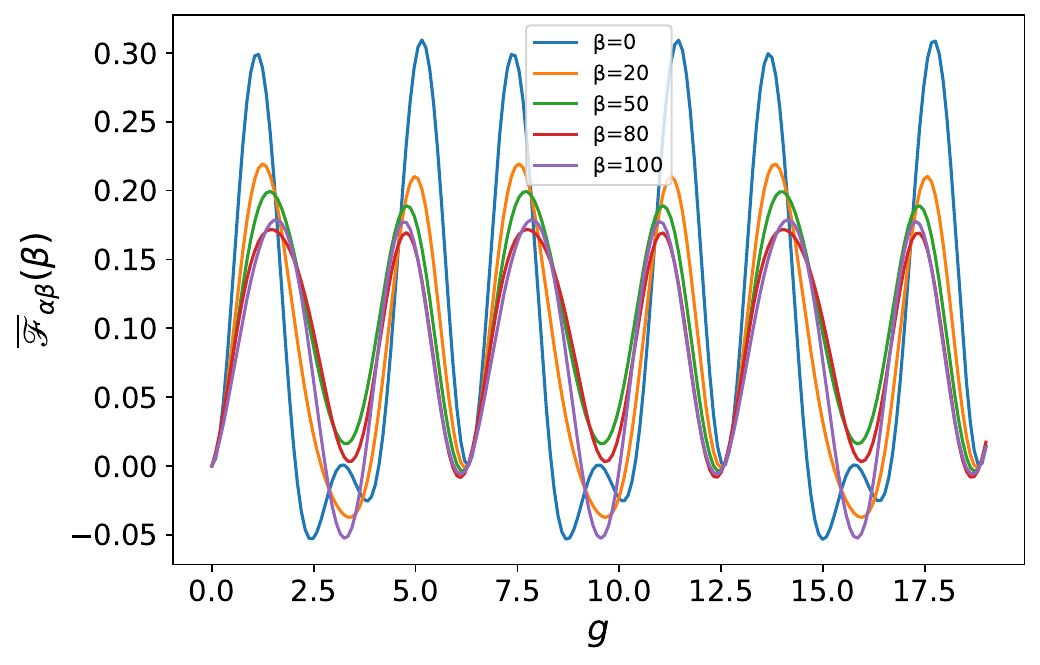}
    \includegraphics[width=0.49\linewidth]{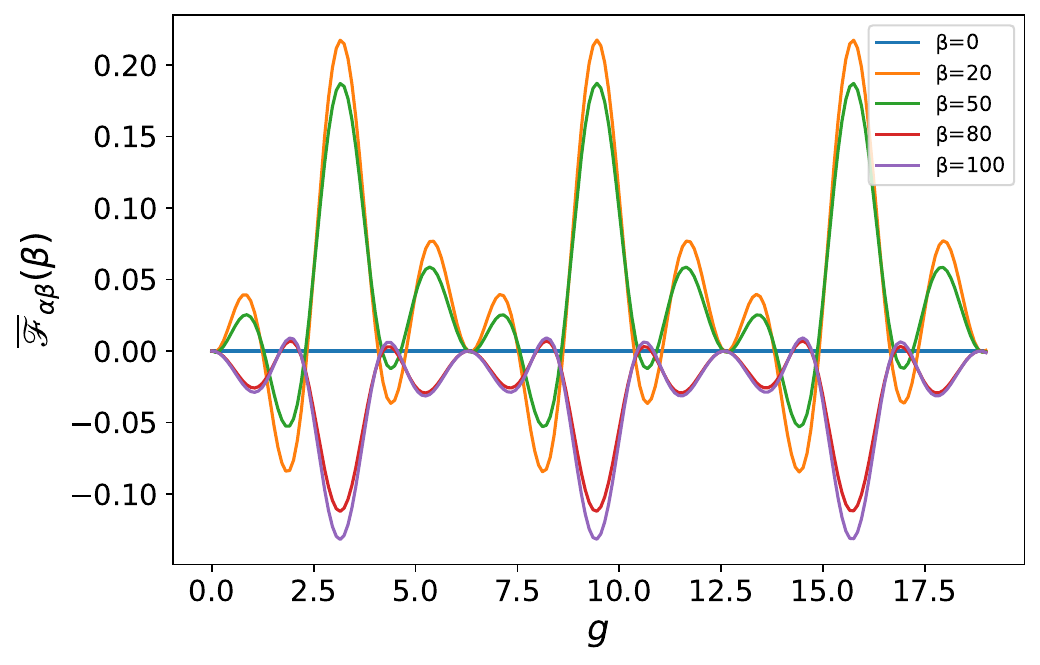}
    \caption{\textbf{Left:} Arbitrary state averaged fidelity for $1^{\rm st}$ qubit swap ($\Delta^{(0,1)}$) as a function of Coupling strength $g$ averaged over 100 iterations ($n_s =100$). \textbf{Right:} Arbitrary state averaged fidelity for $2^{\rm nd}$ qubit swap ($\Delta^{(0,2)}$) as a function of Coupling strength $g$ averaged over 100 iterations ($n_s =100$).}
    \label{neofidelity}
\end{figure}
This theoretical argument is directly and strikingly confirmed by our numerical simulations. We explicitly calculated the average state fidelity $\overline{\mathfrak{F}}_{\alpha\beta}$ for an arbitrary input state (Fig.~(\ref{neofidelity})) and compared it to the basis-state signal $\mathfrak{F}_{Z}$ from Figs.~(\ref{sq1},\ref{sq2}). The comparison demonstrates that the arbitrary-state averaged fidelity provides no new qualitative information beyond the basis-state signal. The fidelity structure for $\Delta^{(0,1)}$ and $\Delta^{(0,2)}$ swap operators shows the exact same periodic oscillation qualitatively and the same monotonic decay of peak fidelity as $\beta$ increases. Quantitatively, in both the swap operator cases, the arbitrary state fidelity is a scaled-down echo of the basis state fidelity. This qualitative match proves that the arbitrary state averaged fidelity $\overline{\mathfrak{F}}_{\alpha\beta}$ is entirely governed by the ``classical" transmission of the basis state. The quantum coherence terms( off-diagonal terms) are indeed suppressed, just as the random phase approximation predicts.\\
Given this mathematical formalism and direct numerical evidence, it is clear that using an arbitrary input state for the single-qubit case unnecessarily complicates the analysis of the fundamental traversability mechanism while yielding no additional physical insights. The chaotic protocol itself effectively decoheres the quantum phase, meaning the fidelity for any input state simply collapses to a measure of the basis-state transmission. Following standard methodology used in theoretical works and recent experimental realizations, we effectively set the arbitrary state $\alpha\ket{0}+\beta\ket{1}$ to a $\ket{0}$ state. This allows our fidelity metric defined in Eq.~(\ref{sqfide}), to serve as a direct and robust teleportation signal for wormhole traversability. We then provide the definitive demonstration of the channel's capacity to transmit genuine quantum information and preserve entanglement in Sec.~(\ref{bell}), where we analyze the teleportation of a maximally entangled Bell state using Pauli-stabilizer-based fidelity.
\section{}
\label{appendix1}
Expanding further on Eq.(\ref{fidestab}), we now find fidelity for an ancilla dual qubit channel carrying $\Phi^{+}$ Bell state message. This is essential to check the consistency of our newly defined Pauli stabilizer-based fidelity approach. consider a dual qubit channel carrying our $\Phi^{+}$ state.

\begin{figure} [h]
    \centering
\begin{tikzpicture}
  \draw[thick] (0,0) -- (5,0) node[right] {$q_0$};
  \draw[thick] (0,-1) -- (5,-1) node[right] {$q_1$};

  \node at (2.5, 0.5) {\large Bell State: $\ket{\Phi^+} = \frac{1}{\sqrt{2}}(\ket{00} + \ket{11})$};

  \draw [decorate,decoration={brace,amplitude=10pt,mirror}] (5.1,0.1) -- (5.1,-1.1) node[midway,xshift=15pt] {\footnotesize entangled};

  \draw[->, thick, blue] (1,0) -- (2,0);
  \draw[->, thick, blue] (1,-1) -- (2,-1);

  \node[draw,rounded corners,fill=gray!20] at (0, -0.5) {Bell Source};
\end{tikzpicture}

    \caption{Two Ancilla qubits carrying $\ket{\Phi^{+}}$}
    \label{fig:enter-label}
\end{figure}
Since there is no Unitary or any other kind of operation on the qubits, the information in the qubit remains unchanged, and so will the overlap. Ideally, we should achieve perfect fidelity in this case, as the initial state will be the same as the final state. From definitions of $S_1,S_2,S_3$ we have
\begin{equation}
    \expval{S_1} = \bra{\Phi^+}X \otimes X\ket{\Phi^+}, \expval{S_2} = \bra{\Phi^+}Z \otimes Z\ket{\Phi^+},\expval{S_3} = \bra{\Phi^+}Y \otimes Y\ket{\Phi^+}
\end{equation}
These equations yield 
\begin{equation}
    \expval{S_1} = 1, \expval{S_2} =1, \expval{S_3} = -1
\end{equation}
Plugging these values back in Eq.(\ref{fidestab}) we have
\begin{equation}
    \mathcal{F}_{\Phi^{+}} = \frac{1}{2}(1+1+1-1)= 1
\end{equation}
Similarly, for Other triplet Bell states, we can see that 
\begin{equation}
     \expval{S_1} = \bra{\Phi^-}X \otimes X\ket{\Phi^-}, \expval{S_2} = \bra{\Phi^-}Z \otimes Z\ket{\Phi^-},\expval{S_3} = \bra{\Phi^-}Y \otimes Y\ket{\Phi^-}
\end{equation}
\begin{equation} \label{A5}
    (\Phi^-) :\expval{S_1} = -1, \expval{S_2} =1, \expval{S_3} = 1
\end{equation}

\begin{equation}
     \expval{S_1} = \bra{\Psi^+}X \otimes X\ket{\Psi^+}, \expval{S_2} = \bra{\Psi^+}Z \otimes Z\ket{\Psi^+},\expval{S_3} = \bra{\Psi^+}Y \otimes Y\ket{\Psi^+}
\end{equation}
\begin{equation} \label{A7}
    (\Psi^+): \expval{S_1} = 1, \expval{S_2} =-1, \expval{S_3} = 1
\end{equation}
From values obtained in Eq.~(\ref{A5}, \ref{A7}), $\mathcal{F}_{\Phi^{-}}=\mathcal{F}_{\Psi^{+}} = 1$ for both these triplet states.
However, for singlet state we get $\mathcal{F}_{\Psi^{-}} = -1$ as 
\begin{equation}
     (\Psi^-): \expval{S_1} = -1, \expval{S_2} =-1, \expval{S_3} = -1
\end{equation}
 For the singlet state, this formula yields an unusable value of fidelity because the singlet state is asymmetric. Thus, we can conclude that for symmetric Bell states ($\Phi^{+},\Phi^{-},\Psi^{+}$), this stabilizer-based definition of fidelity can be explicitly used in the context of wormhole-inspired teleportation protocol.

   \end{widetext} 
\end{document}